\documentclass[twocolumn]{aastex62}
\usepackage{amsmath,amstext}
\usepackage{multirow}
\usepackage{booktabs}
\usepackage[T1]{fontenc}
\usepackage{apjfonts} 

\def\kms{km~s$^{-1}$}
\def\cm2{cm$^{-2}$}

\shorttitle{Metal-Bearing Halo Gas at $z\sim2$}
\shortauthors{Rudie et al.}

\begin{document}

\title{\large \textbf{The Column Density, Kinematics, and Thermal State of Metal-Bearing Gas within \\the Virial Radius of $z\sim2$ Star-Forming Galaxies in the Keck Baryonic Structure Survey}}

\author{Gwen C. Rudie}
\affiliation{The Observatories of the Carnegie Institution for Science, 813 Santa Barbara Street, Pasadena, CA 91101, USA} 
\correspondingauthor{Gwen C. Rudie}\email{gwen@carnegiescience.edu}
\author{Charles C. Steidel}
\affiliation{Cahill Center for Astronomy and Astrophysics, California Institute of Technology, MS 249-17, Pasadena, CA 91125, USA}
\author{Max Pettini}
\affiliation{Institute of Astronomy, Madingley Road, Cambridge CB3 OHA, UK}
\author{Ryan F. Trainor}
\affiliation{Department of Physics \& Astronomy, Franklin \& Marshall College, 415 Harrisburg Pike, Lancaster, PA 17603, USA}
\author{Allison L. Strom}
\affiliation{The Observatories of the Carnegie Institution for Science, 813 Santa Barbara Street, Pasadena, CA 91101, USA}
\author{Cameron B. Hummels}
\affiliation{Cahill Center for Astronomy and Astrophysics, California Institute of Technology, MS 249-17, Pasadena, CA 91125, USA}
\author{Naveen A. Reddy}
\affiliation{Department of Physics and Astronomy, University of California, Riverside, 900 University Avenue, Riverside, CA 92521, USA}
\author{Alice E. Shapley}
\affiliation{Department of Physics \& Astronomy, University of California, Los Angeles, 430 Portola Plaza, Los Angeles, CA 90095, USA}

\begin{abstract}
We present results from the Keck Baryonic Structure Survey (KBSS) including the first detailed measurements of the column densities, kinematics, and internal energy of metal-bearing gas within the virial radius (35-100 physical kpc) of eight $\sim L^*$ galaxies at $z\sim2$. From our full sample of 130 metal-bearing absorbers, we infer that halo gas is kinematically complex when viewed in singly, doubly, and triply ionized species. Broad \ion{O}{6} and \ion{C}{4} absorbers are detected at similar velocities to the lower-ionization gas but with very different kinematic structure indicating that the circumgalactic medium (CGM) is multi-phase. There is a high covering fraction of metal-bearing gas within 100 kpc including highly ionized gas such as \ion{O}{6}; however, observations of a single galaxy probed by a lensed background QSO suggest the size of metal-bearing clouds is small ($<400$ pc for all but the \ion{O}{6}-bearing gas).
The mass in metals found within the halo is substantial, equivalent to $\gtrsim 25\%$ of the metal mass within the interstellar medium. The gas kinematics unambiguously show that 70\% of galaxies with detected metal absorption have some unbound metal-enriched gas, suggesting galactic winds may commonly eject gas from halos at $z\sim2$. Significant thermal broadening is detected in CGM absorbers which dominates the internal energy of the gas. 40\% of the detected gas has temperatures in the range $10^{4.5-5.5}$ K where cooling times are short, suggesting the CGM is dynamic, with constant heating and/or cooling to produce this short-lived thermal phase. 
\end{abstract}


\section{Introduction}

The gaseous environments of galaxies are a crucial but poorly constrained component of galaxy formation and evolution. The circumgalactic medium (CGM) is the principal reservoir for future gas accretion, and its kinematics, thermal properties, and metal enrichment provide vital constraints on the properties of cosmological inflow and galaxy-scale outflows. Characterization of this gas therefore presents a unique window into the baryonic flows that are expected to profoundly influence the formation and evolution of galaxies.

The last decade has seen considerable growth in observations of the low-redshift CGM, in large part due to the installation of the Cosmic Origins Spectrograph on the Hubble Space Telescope. These observations have demonstrated that low-redshift galaxies are surrounded by substantial reservoirs of metal-enriched gas \citep{tum11,wer14,lia14,joh17,zah18} which correlates with the environments in which the galaxies reside \citep{joh15,bur16,nie18} and the mass and star-formation rate of the system  \citep{che10,tum11,bor13,bor14,joh17,rub18}. At low redshift there is evidence that strong metal absorbers are more commonly found along the major and minor axis of disk galaxies \citep{kac12}, suggestive of metal-bearing gas accreting along the major axis disk while metal-enriched outflows escape along the minor axis.  

While our understanding of the low-$z$ CGM is growing, arguably the best epoch in which to study the effect of gas flows on galaxy formation is $2<z<3$, during the peak of cosmic star formation \citep{mad96} and super-massive black hole growth \citep{ric06}.  At these redshifts, spectroscopic observations of star-forming galaxies commonly exhibit signatures of strong baryonic outflows \citep{pet01, sha03, rup18}. At the same time, the baryonic accretion rate onto galaxies is predicted to be near its maximum (e.g., \citealt{fau11b, van11b, red12b}). 
Therefore, we expect the signatures of baryonic flows to be most readily observable at $2<z<3$. Absorption line spectroscopy of bright background sources provides a uniquely sensitive probe of the physical properties and chemical compositions of gas in close proximity to galaxies.

In fact, the strong rest-UV transitions most useful for characterizing cosmic gas are more accessible in the distant universe than at low redshift. Thanks to cosmological expansion, these transitions fall in the observed optical, allowing observations with large-aperture ground-based telescopes equipped with high-dispersion spectrographs. For this reason, one can obtain higher-resolution and higher signal-to-noise-ratio CGM data for $z>2$ galaxies than can be obtained for low-redshift galaxies. 

Nevertheless, observational studies of the high-redshift CGM require significant observational investments due to the extreme faintness of typical galaxies in the distant universe, complicating the assembly of large samples of galaxy-QSO pairs. Thus, only a handful of studies at $z\gtrsim2$ have been possible to date \citep{ade03,ade05,hen06,cri11,gcr12,rub15}. 

The first CGM studies at $z>2$ were carried out by \citet{ade03} and \citet{ade05} who studied \ion{H}{1} and \ion{C}{4} absorption surrounding over 1000 $z>2$ Lyman Break Galaxies (LBGs).\footnote{LBGs are typical star-forming galaxies in the distant universe, selected based on their blue UV continua and/or a strong break at 912\AA\ in their rest frame due to \ion{H}{1} Lyman continuum absorption from the Lyman alpha forest along their line of sight.} These authors found strong correlations between the locations of galaxies and \ion{H}{1} and \ion{C}{4} absorption and enhanced \ion{H}{1} absorption on Mpc scales. In this work, very strong correlations are reported between \ion{C}{4} absorbers and galaxies, suggesting strong \ion{C}{4} absorbers and LBGs must be residing in the same halos. 

\citet{sim06} further analyzed a subset of these data, considering detailed multi-ion column density measurements from line fitting for 10 galaxies within 500 pkpc of the line of sight to HS1700$+$6416. These authors also calculated photionization corrections to determine the sizes and metallicities of gas within the CGM. Their results suggest that metal absorbers within the CGM of the galaxies are metal-rich ($>0.1$ Z$_\odot$) and small ($<1$ kpc) implying relatively high gas densities (1000 times the mean density at the redshift of the galaxies).

\citet{ccs10} stacked the spectra of many background galaxies surrounding $\sim500$ foreground $z>2$ LBGs to study the CGM at very close impact parameters (<125 pkpc). They found a high covering fraction and large equivalent width of \ion{H}{1} and many metal species that declined as the impact parameter increased. Past $\sim60$ pkpc, ionic metal line absorption is no longer detected in these stacks; and so to probe the CGM at slightly higher impact parameters, a higher fidelity measurement of the CGM absorption, such as that provided by background QSOs, is needed. 

More recently, higher S/N QSO spectra have been combined with very large redshift surveys in the Keck Baryonic Structure Survey (KBSS; \citealt{gcr12}). \citet{rak12} presented the first 2D maps of \ion{H}{1} absorption around high-$z$ galaxies using KBSS data. These maps, measured using the pixel optical depth (POD) technique, showed Mpc-scale enhancement in the \ion{H}{1} optical depth surrounding galaxies. The kinematics of the \ion{H}{1} exhibited clear signatures of gas infall on Mpc scales, and large peculiar velocities, plausibly due to gas outflows, on small scales \citep{rak12,gcr12,tur14}. 

\citet{gcr12} presented a detailed Voigt profile decomposition of all \ion{H}{1} absorbers in the CGM of KBSS galaxies. We found very strong enhancements in the column density $N_{\rm HI}$ close to galaxies, finding typically $N_{\rm HI}\approx 10^{16.5}$ \cm2 within 100 kpc, 3 orders of magnitude higher than was typical at random places in the IGM. We also noted that $N_{\rm HI} \ge 10^{14.5}$ \cm2\ absorbers correlate with the positions of galaxies, suggesting they may trace the CGM. The line widths of the \ion{H}{1} absorbers within 300 kpc were found to be larger than typical of the IGM even when compared at fixed $N_{\rm HI}$, suggesting that gas within the CGM is either more turbulent or hotter. 

\citet{tur14} used pixel optical depth (POD) analysis to study \ion{H}{1} and metal line optical depth as a function of distance to KBSS galaxies, finding strong enhancements in metal line absorption within 180 pkpc and $\pm$240 \kms. \citet{tur15} focused on gas close to the galaxies ($<180$ pkpc), studying the enhancement of metal line absorption, finding enhanced \ion{O}{6} optical depth  at fixed \ion{H}{1}, \ion{C}{4}, and \ion{Si}{4} optical depths extending to $\pm 350$ \kms. \citet{tur15} suggested that the \ion{O}{6} detected at high \ion{H}{1} optical depth could be formed via photoionization; however, the large \ion{O}{6}  enhancement at low \ion{H}{1} optical depth was more consistent with collisionally ionized hot gas with relatively high metallicity ($>0.1$ Z$_\odot$), properties most likely found within galactic winds.

\begin{deluxetable*}{llr@{\hspace{0.3in}}rrrrrrr}
\tablecaption{Absorption within the halo of KBSS galaxies}  
\tablehead{
\multirow{2}{*}{Galaxy} & \multirow{2}{*}{D$_{\rm tran}$ [pkpc]} & \multirow{2}{*}{\#\tablenotemark{a}} & \multicolumn{6}{c}{$\log{(\Sigma N_x)}$ [\cm2]\tablenotemark{b}}\\
\cmidrule(lr){4-9}
\colhead{} & \colhead{} & \colhead{} & \colhead{\ion{Si}{2}} & \colhead{\ion{C}{2}}  & \colhead{\ion{Si}{3}}  &   \colhead{\ion{C}{3}}  & \colhead{\ion{Si}{4}}  & \colhead{\ion{C}{4}}}
\startdata
Q0105-BX90b  &   34.5 & 12 &   <12.48 &  <12.42    &    11.70   &   13.57  &	11.88  &   13.51   \\    
Q1442-MD50a  &   56.5 & 15 &    12.45 &   13.34    &    12.99   &  \nodata & 	13.21  &   14.19   \\      
Q1549-D15    &   56.6 & 15 &   <12.58 &  <12.26    &    12.98   &   14.46  &	13.15  &   14.04   \\     
Q1442-MD50b  &   61.9 & 31 &    13.95 &   14.69    &   >13.56   &  >14.22  &   	14.19  &  >14.63   \\       
Q0142-BX182  &   75.2 & 25 &    13.73 &   14.56    &   >13.57   &  >14.08  &   	13.93  &   14.60   \\      
Q2343-BX551  &   86.9 &  8 &   <12.44 &  \nodata   &    13.27   &  \nodata & 	13.27  &   14.15   \\     
Q0100-BX210  &   89.6 & 24 &   <12.45 &  \nodata   &    12.86   &   14.26  &	12.56  &   14.27   \\      
Q1623-BX432  &   97.3 & 0  &   <12.37 &  <12.60    &	\nodata &  \nodata &   <12.29  &  <12.44   \\   
\enddata
\tablenotetext{a}{The number of metal-line Voigt profile components in the fit.}
\tablenotetext{b}{Total log column density within 1000 \kms}
\label{table_N}
\end{deluxetable*}

In this paper we continue our analysis of the KBSS data set. Here we describe the first detailed measurements of metal line absorption within the halos of eight $z\sim2$ galaxies including the column densities, kinematics, and thermal state of gas within the high-$z$ CGM. Section 2 presents the observational data including more information concerning the galaxy sample. Section 3 discusses the Voigt profile modeling of the QSO absorption line data. In Section 4, we present our measurements of the column densities and covering fractions of metal ions in the KBSS. Section 5 covers the kinematics of metal enriched gas, with Section 6 presenting our measurements of the temperature and turbulence within the CGM. In Section 7, we calculate the implied metal mass in the halo and compare our observations to theoretical models and other observations, discussing implications for our understanding of the CGM. A detailed summary of our results is presented in Section 8.

Throughout this paper we assume a $\Lambda$CDM cosmology with $H_{0} = 70$ \kms\  Mpc$^{-1}$, $\Omega_{\rm m} = 0.3$, and $\Omega_{\Lambda} = 0.7$. All distances are expressed in physical (proper) units unless stated otherwise. We use the abbreviations pkpc and pMpc to indicate physical units.

\begin{figure*}
\center
\includegraphics[width=0.25\textwidth]{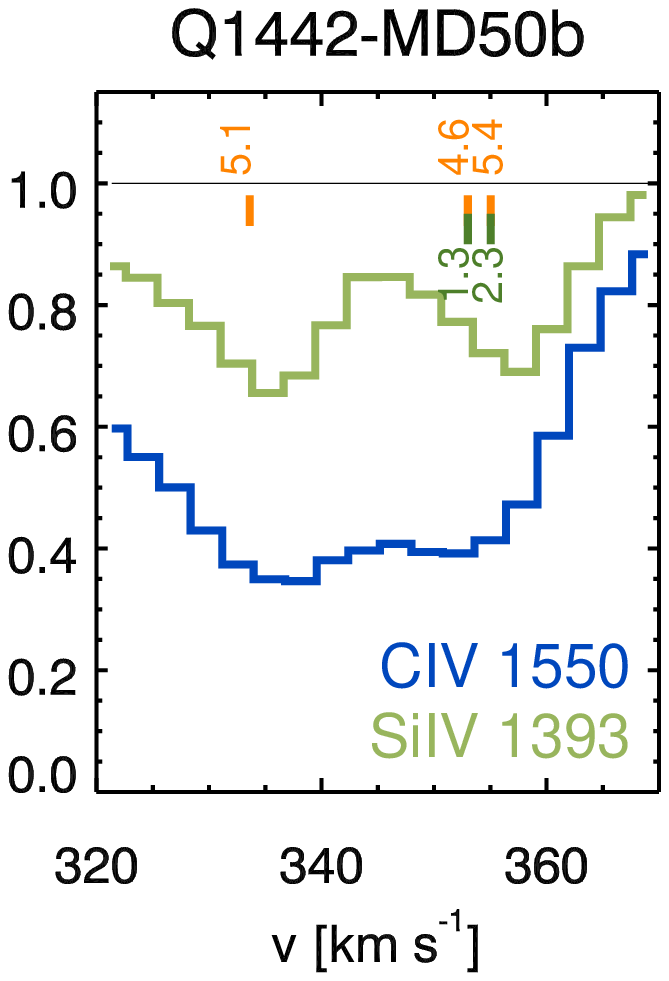}\includegraphics[width=0.25\textwidth]{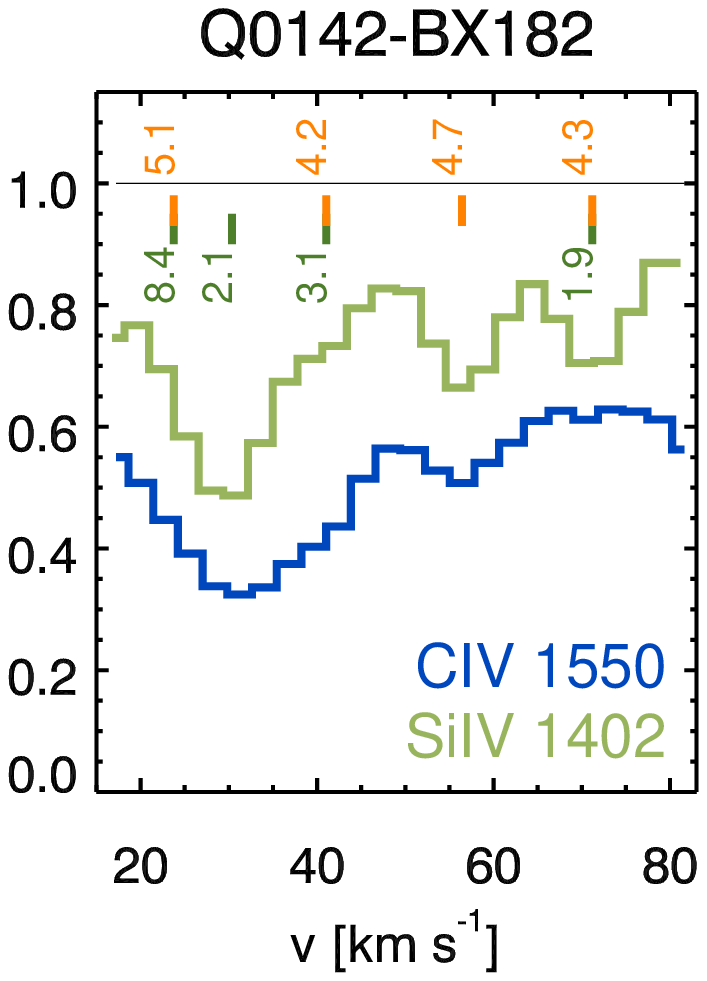}\includegraphics[width=0.25\textwidth]{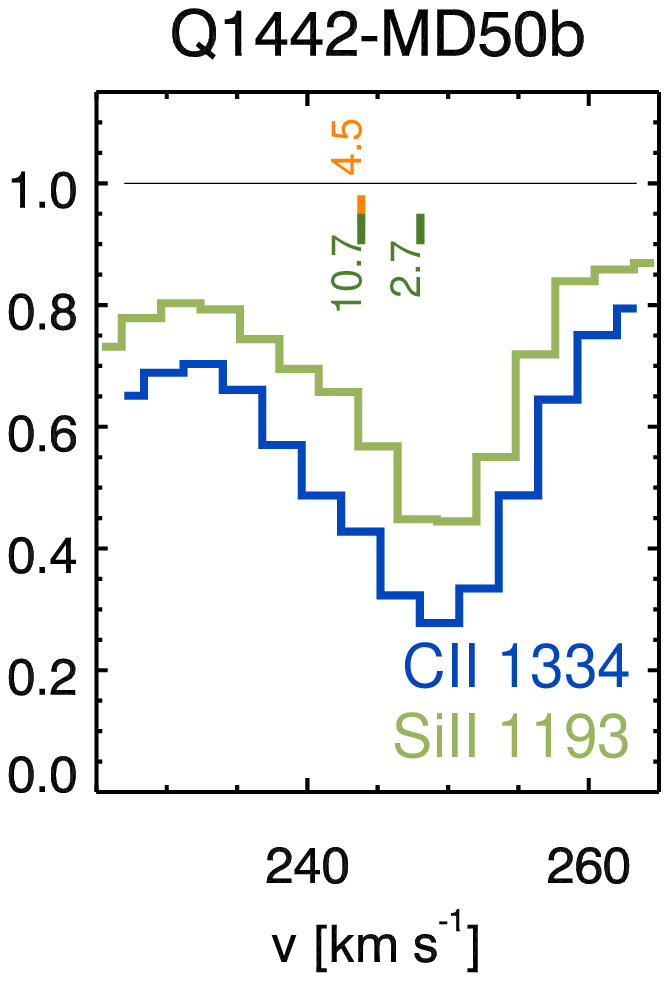}\includegraphics[width=0.25\textwidth]{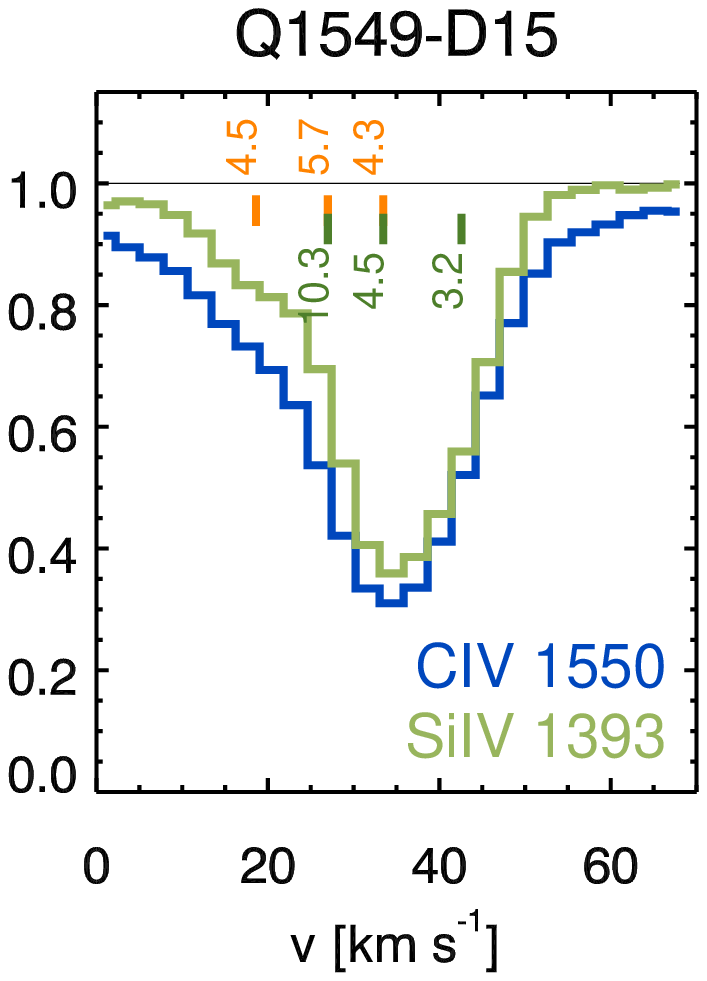}
\caption{Examples of thermally (left two panels) and turbulently (right two panels) broadened absorption systems within the CGM of KBSS galaxies. The blue and green curves are the continuum-normalized QSO spectrum plotted with respect to the systemic redshift of the galaxy. Each panel shows one transition of C (dark blue) and one transition of Si (light green). Because these elements have different atomic masses, the widths of their respective absorption lines can be used to differentiate between thermal and turbulent broadening. Note that each panel compares either triply ionized species or singly ionized species such that ionization differences between the elements is minimized. The two left panels show absorbers where the internal energy of the absorber is dominated by thermal energy. In this case, the widths of the silicon absorption lines are appreciably narrower than those of carbon. The two right-hand panels show cases where the absorbers are dominated by turbulent broadening, in which case the widths of the absorption lines from different mass elements are the same. Tick marks show the location of individual components of the Voigt profile fits to these data. Orange ticks show the locations of thermally broadened lines with inferred temperatures $T>10^4$ K which are labeled with the log of the temperature in Kelvin inferred from the fit. Green ticks are turbulently broadened absorbers with measures $v_{\rm turb}>1$ \kms\ and are labeled with the turbulent velocity inferred from the fit in \kms. Note that some absorbers are found to have little to no detected thermal or turbulent broadening (those with only one tick) while most absorbers are measured to have some level of both thermal and turbulent broadening (those with both thermal and turbulent ticks).}
\label{stack}
\end{figure*}

\section{Observations}

\label{observations}

The data presented here are part of the Keck Baryonic Structure Survey (KBSS; \citealt{gcr12,ccs14,str17}) which combines high resolution spectroscopy of 15 hyperluminous $2.5<z<2.9$ QSOs \citep{tra12} with large rest-optical and rest-UV galaxy redshift surveys focusing on those galaxies in the foreground of the QSO. The background QSO spectroscopy provides high-fidelity absorption line constraints on the gas in proximity to the foreground galaxies. 

The QSO data were taken with the High Resolution Echelle Spectrometer (HIRES; \citealt{hires}) on the Keck I telescope. All publicly available echelle spectroscopy of the QSOs was also incorporated, including both HIRES data from the Keck Observatory Archive (KOA) and UVES \citep{UVES} data from the European Southern Observatory's Science Archive Facility. A detailed description of the QSO data and their reduction is provided in \citet{gcr12}. The final combined QSO spectra have $R\simeq 45,000$ (FWHM$\simeq 7$ \kms), S/N $\sim 50-250$ per pixel, covering at least the wavelength range 3100 -- 6000\AA\ with no spectral gaps. The reduced QSO data were continuum-normalized using low-order spline interpolation prior to the analysis of the detected absorption lines. 

Foreground galaxies in the KBSS fields were selected based on their rest-frame UV colors using criteria described in \citet{ccs03}, \citet{ade04}, and \citet{ccs04}. Optical and near-infrared (NIR) galaxy spectroscopy was carried out using the Low Resolution Imaging Spectrometer (LRIS-B; \citealt{oke95, ccs04} and the Multi-Object Spectrometer for Infrared Exploration (MOSFIRE; \citealt{MOSFIRE1,MOSFIRE2}), both on the Keck I telescope. A detailed discussion of the LRIS survey can be found in \citet{ccs10} and \citet{gcr12}. The MOSFIRE survey is discussed in \citet{ccs14} and \citet{str17}.

All of the galaxies presented herein have redshifts measured using rest-frame optical spectroscopy with Keck/MOSFIRE. These redshifts are derived from bright nebular emission lines (typically H$\alpha$ and [\ion{O}{3}]$\lambda5007$) and have associated errors $\lesssim 10$ \kms~\citep{ccs14}. Impact parameters between the QSO line of sight and the galaxy are calculated using the measured angular separation of the QSO and galaxy, and the galaxy redshifts determined from the rest-optical spectra.\footnote{One of the QSOs in our sample (Q0142) is lensed (see Section \ref{size}), and so we calculate the distance between the galaxy and the location of the brightest image at the redshift of the galaxy. The two images are only 400 pc separated from each other at the redshift of the galaxy.} For this calculation, we assume that the galaxy redshifts are purely cosmological. 

In this work, we present analysis of gas within 100 pkpc, or roughly the virial radius \citep{tra12,rak13,tur17}, of galaxies in the KBSS sample. All the galaxies included in this analysis have QSO spectroscopy covering at least two Lyman series transitions so that the \ion{H}{1} column is well constrained. While not critical for the analysis presented herein, the well-constrained \ion{H}{1} column densities will allow for measurements of gas phase metallicities in future work. The names of the galaxies and the impact parameter to the line of sight to the QSO are listed in Table \ref{table_N}. The KBSS galaxy sample probed within $R_{\rm vir}$ comprises 8 galaxies and is the first published sample of high-redshift CGM constraints on these scales. The absorption within the KBSS high-$z$ halos is complex, providing a statistical sample of 130 metal-bearing absorbers. Future work will present metal line analysis at larger distances for the full KBSS sample. 

The $2.1<z<2.7$ galaxies considered in this paper (aside from Q0105-BX90b, see below) have stellar masses $9<\log{( M_*/ M_\odot)}<10.7$ and star-formation rates of $7<$SFR$<52$~M$_\odot$~yr$^{-1}$ determined through spectral energy distribution fits to their broad-band photometry \citep{str17}. The galaxies are drawn from a representative sample of typical star-forming galaxies at $z\sim2$ characterized by the following properties. The parent sample of galaxies are young ($\sim 50$ Myr - 1 Gyr, \citealt{erb06c,red12a,the18}) and gas-rich \citep{erb06c,tac10}, with gas-phase oxygen abundances $12+\log{(\rm O/H)}=8.37$, $Z=0.5Z_\odot$ \citep{str18}. Due to their young ages and high star-formation rates, they exhibit comparatively Fe-poor stellar populations which have a 1-4 Ryd ionizing spectrum  that is harder than is typical in the local universe, but likely common at high redshift \citep{ccs14,str17,ccs16}. Clustering analysis of the parent sample suggests these galaxies occupy dark matter halos with median halo masses $M_{\rm halo}=10^{11.9} M_\odot$ \citep{tra12,rak13,tur17} corresponding to a typical virial radius of $\sim100$ pkpc at these redshifts. 

The closest galaxy in the sample, Q0105-BX90b, was serendipitously found during MOSFIRE spectroscopic observations of another UV-color selected galaxy. BX90b lies only 4\arcsec\  from the QSO line of sight, where the wings of the point spread function from the bright QSO make accurate photometry of faint sources difficult. Due to this and the  faintness of the galaxy, its full properties are hard to quantify accurately with the photometry we have in hand. We estimate that BX90b is at least 1 magnitude fainter than typical KBSS galaxies ( ${\cal R}\sim26$;  $K_s \sim$ 25), which would also place it at or below the low-mass end of the KBSS stellar mass distribution, i.e., $M_*<10^9 ~M_\odot$.

\section{Analysis of the QSO Absorption Spectra}

\label{fits}

In this section, we describe the fitting methodology used to measure the redshifts, column densities, and line widths of metal absorbers within 100 pkpc of galaxies in the KBSS survey. We model absorbers in the QSO spectra using a series of Voigt profiles, allowing us to constrain the bulk and internal kinematics of the gaseous structures.

Section \ref{temp_method} explains how thermal and non-thermal broadening can be measured using this data.. Section \ref{multiphase} describes observational evidence for two phases of absorbing gas, one which appears in all species with ionization potentials $<48$ eV, and a separate \ion{O}{6}-bearing phase. We model these two phases separately as described in Sections \ref{vpfit} and \ref{O6}, respectively.

\subsection{Thermal and Turbulent Broadening of Absorbers}

\label{temp_method}

In this work, we present the first measurements of the temperature and turbulent velocities of gas in the high-$z$ CGM. These measurements are enabled by the very high S/N and high resolution of our data ($R\simeq 45,000$, FWHM$\simeq 7$ \kms) which resolves the internal broadening of absorber sub-components. 
For single ionization stages, one measures the absorption line width, parameterized in Voigt profiles as the Doppler width, $b_{\rm d}=\sqrt{2}\sigma$ where $\sigma$ is the one-dimensional velocity dispersion of the gas (along the line of sight), which depends on the combined effect of all motions of the atoms or ions in the gas cloud.
There are several possible physical scenarios that lead to the broadening of IGM and CGM absorbers. Most relevant to this work are the effects of gas temperature and bulk motions of the gas (hereafter referred to as turbulence).\footnote{Another plausible source of broadening for the most physically extended absorbers with sizes $>50-100$ kpc is differential Hubble flow across gas ``clouds'' (see \citealt{gcr12b}). This source of broadening is likely only common in low-$N_{\rm HI}$ absorbers.}

For an isothermal gas cloud, the observed thermal broadening depends on the temperature of the gas, $T$, and the mass of the ion, $m$: 
\begin{equation}
\label{b_therm}
b_{\rm therm}^2 = \frac{2kT}{m}
\end{equation}
where $b_{\rm therm}$ is the thermal component of the Doppler parameter and $k$ is the Boltzmann constant. For non-thermal broadening such as turbulence, the velocity within the gas is the same for all species and does not depend on the masses of the ions. 

When transitions from multiple elements are observed arising from the same gaseous structure, it is possible to decompose the broadening of each absorber into a thermal and non-thermal component. For example, in the case of absorbers detected in \ion{C}{4} and \ion{Si}{4}:
\begin{equation}
\label{temp_turb1}
b_{\rm CIV}^2 = b_{\rm turb}^2 + b_{\rm CIV,~therm}^2 = v_{\rm turb}^2 + \frac{2kT}{m_{\rm C}}
\end{equation}
\begin{equation}
\label{temp_turb2}
b_{\rm SiIV}^2 = b_{\rm turb}^2 + b_{\rm SiIV,~therm}^2 = v_{\rm turb}^2 + \frac{2kT}{m_{\rm Si}}
\end{equation}
where $v_{\rm turb}$ is the turbulent velocity.
Because the atomic masses of Si and C differ by a factor of 2.3, with measurements of $b_{\rm CIV}$ and $b_{\rm SiIV}$ it is possible to solve for $T$ and $v_{\rm turb}$.

In practice, for typical gas temperatures in the IGM of $2 \times 10^4$ K \citep{hui97,sch99, gcr12b}, the thermal broadening for C absorbers is $\sim 5$ \kms\ and for Si absorbers is $\sim 3$ \kms, which are at the resolution limit of our data; however, for higher gas temperatures and turbulent velocities, which appear to be common in the CGM at $z\sim2$, the broadening can be more easily detected. Figure \ref{stack} shows examples of several absorption complexes in the CGM of KBSS galaxies. The left two panels show clear evidence of thermal broadening: the heavier ion (Si, shown in light green) has narrower line widths than that of the lighter element (C, shown in dark blue). The right two panels show CGM absorbers where the silicon and carbon profiles appear quite similar, which requires that thermal broadening be sub-dominant.

As we describe below, we use the joint constraint provided by multiple elements (typically carbon and silicon) to constrain the thermal and turbulent energy of the gas. While in principle absorbers detected in one metal ion as well as \ion{H}{1} can also be used to constrain $T$ and $v_{\rm turb}$, in the $z\sim2.3$ CGM most \ion{H}{1} absorbers have multiple metal absorption components detected within their velocity range, making their association ambiguous (the thermal broadening of \ion{H}{1} in $2\times10^4$ K gas is $b_{\rm HI,~therm}=18$ \kms). For this reason, we do not compare the  Doppler widths of \ion{H}{1} with those of metal transitions to constrain the gas temperature or turbulence. 

We caution that while the quantities $T$ and $v_{\rm turb}$ are separable in the way described above, uncertainties in the Doppler widths propagate into the thermal and turbulent broadening measurements causing the two quantities to be correlated. The detection of additional elements, particularly more massive elements (such as iron), would improve our ability to disentangle the two quantities. However in practice, Fe absorption is rarely detected in the CGM of these galaxies. Section \ref{vpfit} describes in detail how we measure $T$ and $v_{\rm turb}$ in this sample, and Section \ref{temp} discusses the results of these measurements. 

\begin{figure*}
\center
\includegraphics[width=0.45\textwidth]{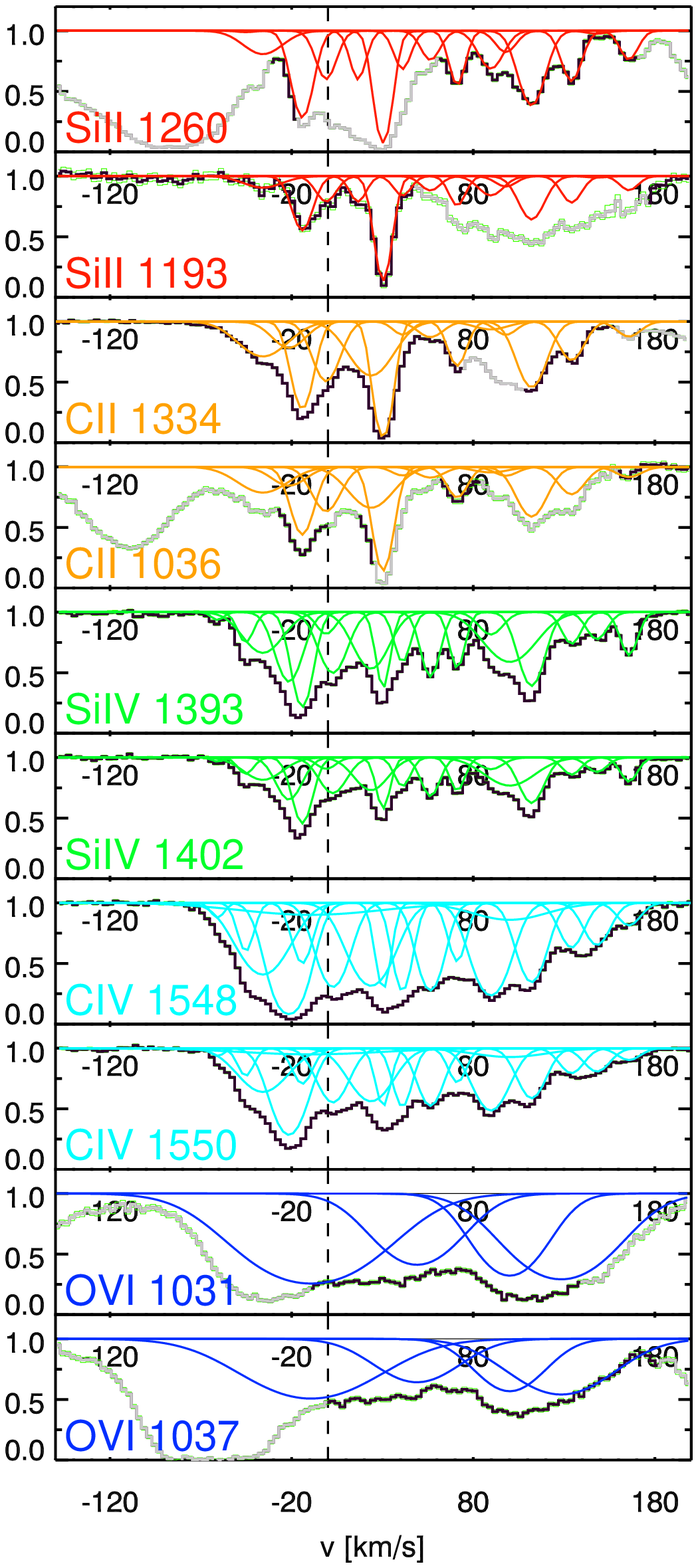} \includegraphics[width=0.45\textwidth]{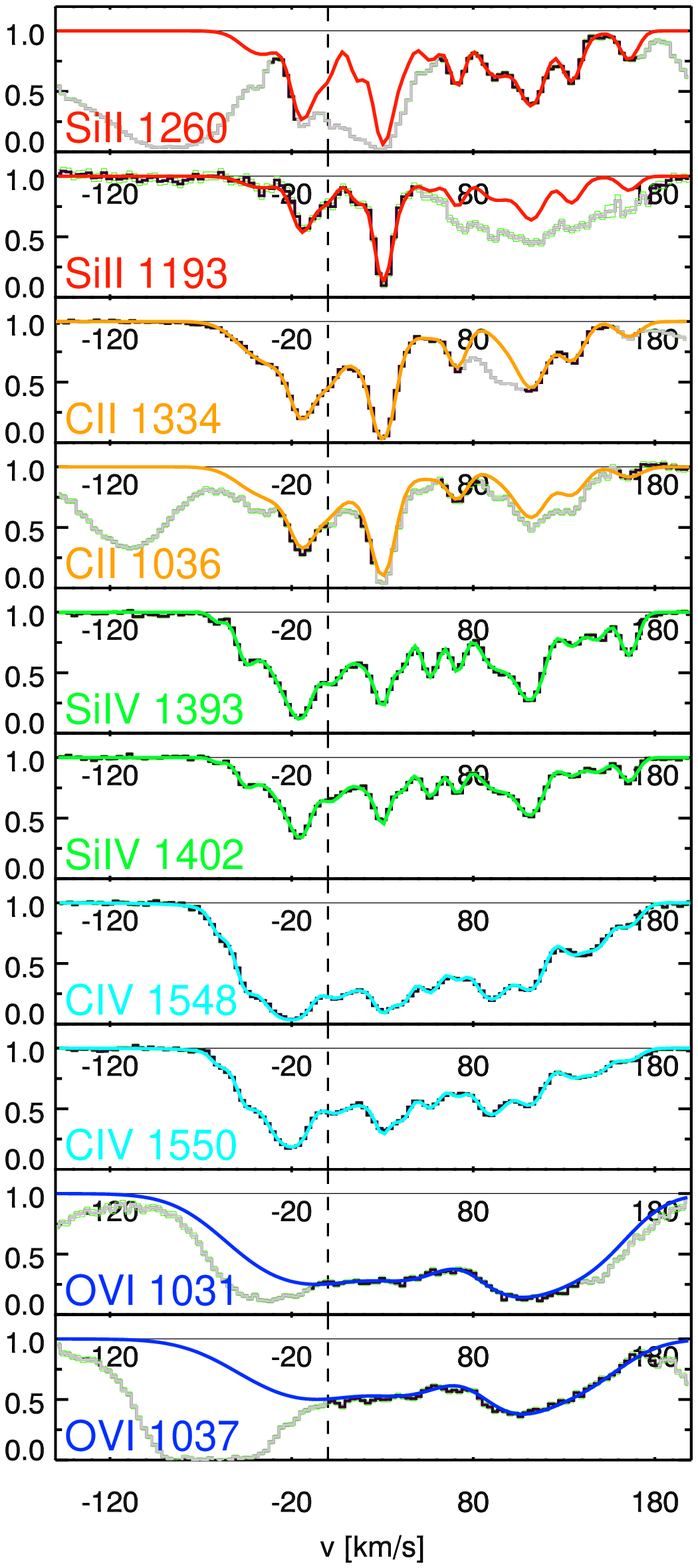}
\caption{Example of the complex multi-component absorption structure within the halo of Q0142-BX182, one of the most complex systems in the KBSS sample. The velocity scale is given with respect to the systemic redshift of the galaxy, measured based on strong rest-frame optical emission lines (H$\alpha$, [\ion{O}{3}]). The continuum normalized QSO spectrum is shown in black. The green curves that trace the black QSO spectrum with little to no deviation indicate the $\pm1\sigma$ error on the normalized flux. Grey sections of the QSO spectrum are highlighted to show regions where there is contamination from \ion{H}{1} or other metal ions within the wavelength range fit. Colored curves show the Voigt profile decomposition of the absorption. The left-hand panel shows the individual component structure while the right-hand panel shows their product, which can be used to compare directly to the data to test the goodness of fit. Note that strong absorption is detected in multiple ionization phases. \ion{C}{3} and \ion{Si}{3} absorption is also detected (not shown) but is saturated and so is not useful for comparing the subcomponent structure. Note that the same pattern of absorption features is present in the low and high-ion gas, but not in \ion{O}{6}. Some of the broad features seen in \ion{O}{6} do appear in the fit to \ion{C}{4}. We suggest the the gas observed in \ion{Si}{2} - \ion{C}{4} represents a single phase while the broad features detected in \ion{O}{6} and \ion{C}{4} represent a second phase.}
\label{BX182}
\end{figure*}

\subsection{Observational Evidence that CGM is Multiphase}

\label{multiphase}

Visual inspection of the data suggests that low-, intermediate-, and high-ionization species with ionization potentials less than 48 eV (equal to or lower than that of \ion{C}{4}) typically share the same sub-component structure where detected (see Figure \ref{BX182}). While the kinematic structure in the gas is often complex, the similarity of all of these components suggests that the low, intermediate, and high ions (excluding \ion{N}{5} and \ion{O}{6}, see Section \ref{O6}) are co-spatial within the CGM. For this reason, each detected sub-component (absorption modeled with a single Voigt profile) is presumed to exist in all of the ionization stages of the various metal species with ionization potentials less than 48 eV (i.e. \ion{C}{2}, \ion{C}{3}, \ion{C}{4}, \ion{Si}{2}, \ion{Si}{3}, \ion{Si}{4}, etc.) 

Conversely \ion{O}{6} (and the rarely detected \ion{N}{5}) do not appear to have the same kinematic structure as the lower-ionization gas. When the pathlength corresponding to \ion{O}{6} absorption is free of contamination, the observed \ion{O}{6} is found at the same velocities as the lower ionization gas, but the \ion{O}{6} is distributed in a small number of broad absorption systems, suggestive of a very different origin than the lower-ionization gas (as has been previously reported in the high-$z$ IGM, \citealt{sim02}.) 

Given the very different spectral morphology of the \ion{O}{6}-bearing gas from that seen in lower-ionization potential species, the data support the view of the CGM as multi-phase. In this scenario, the gas giving rise to \ion{O}{6} absorption has potentially very different densities and temperatures than the lower-ionization gas. 

As such, \ion{O}{6} and \ion{N}{5} are not assumed to have the same component structure as the lower-ionization species. These broad features, which are also often detected in \ion{C}{4} (see Figure \ref{BX182}), are modeled with multiple ionic transitions as well; however, the choice of which ions are included is motivated by both the data and by collisional and photoionization non-equilibrium models \citep{gna17,opp13} as described in Section \ref{O6}.

\subsection{Voigt Profile Fitting}

\label{vpfit} 

In order to model the absorption signal due to gas in the halos of the KBSS galaxies, we decompose the absorption into a series of Voigt profiles. Each absorber complex contains a variable number of sub-components. Each component is modeled with a single redshift, and a single value for the gas temperature and turbulent broadening of the absorber, but with column densities that vary for each unique ionization state of each element. In practice, most commonly, we detect some subset of \ion{C}{2}, \ion{C}{3}, \ion{C}{4}, \ion{Si}{2}, \ion{Si}{3}, \ion{Si}{4}, and \ion{O}{6}, with other ions detected less frequently. In the case where the measured temperature of the gas in an individual sub-component is too hot to expect low-ionization stages to exist (see Section \ref{O6}), we model that sub-component  with high-ionization species only. 

In order to measure the parameters of absorbers in the CGM of KBSS galaxies, we employ  VPFIT\footnote{In this work, we run version 9.5 of VPFIT. VPFIT is available at https://www.ast.cam.ac.uk/$\sim$rfc/vpfit.html} written by R. F. Carswell and J. K. Webb. VPFIT is a $\chi^2$ minimization Voigt profile fitting program designed to fit complex blends of multiple absorption components in order to derive the physical parameters of the gas. The output of VPFIT includes the measured column densities ($N$), Doppler widths ($b_{ \rm d}$), and redshifts of absorbers as well as errors on all of the above parameters. It is also possible to tie the constraints on the Doppler widths of multiple elements in the same component in order to constrain $T$ and $v_{\rm turb}$ directly. In order to derive meaningful fits to the complex absorption systems associated with galaxies, detailed initial estimates of the component structure and parameters of each absorber are required. 

The process of locating metal line absorption in the KBSS QSO spectra is significantly aided by the high signal-to-noise ratio of our QSO data. Typical KBSS QSO spectra have S/N $\sim$ 100 in the regions redward of the Ly$\alpha$ forest where many metal lines are detected. Due to the high S/N of the data, deviations in the continuum are nearly always true absorption systems, so a visual inspection of the QSO spectra in the regions of interest can identify the locations of metal absorption systems. However, given the complexity of some of the metal systems, we prefer to automate the initial fitting of absorbers. 

For the initial fit, we typically use the \ion{C}{4} section of the spectrum to make  estimates of the location and component structure of the absorbers. We employ \ion{C}{4} absorbers because their spectral range is mostly free of contamination, and because \ion{C}{4} absorption is strong in these systems.\footnote{For one galaxy in the sample, Q0142-BX182 , the absorption component structure is sufficiently blended in \ion{C}{4} that we opt instead to use the \ion{Si}{4} section of the spectrum.} An automated search using custom-built code is conducted within regions of the spectrum corresponding to \ion{C}{4} absorption within $\pm1400$ \kms\ of the systemic redshift of galaxies.\footnote{We use a $\pm1400$ \kms\ as a conservative upper limit on the velocity based on the results of \ion{H}{1} studies of the CGM of KBSS galaxies \citep{gcr12}.} Absorbers are located through cross-correlation of a representative template absorber (a single non-saturated Gaussian absorber) with the HIRES spectrum. Peaks in the cross-correlation are presumed to be the locations of individual absorption components which are subsequently fit with a Gaussian to estimate their widths and column densities. This method is analogous to that used to make initial estimates for \ion{H}{1} absorbers in the spectra as described in \citet{gcr12}.

Once initial estimates of the parameters of \ion{C}{4} absorbers are complete, the parameters are input into VPFIT. The output of VPFIT is checked by eye, and the number of components and their parameters are adjusted and iteratively rerun until an appropriate fit to the \ion{C}{4} absorbers is achieved (reduced $\chi^2 \approx 1$.)

Once an appropriate fit to \ion{C}{4} is complete, the parameters of the \ion{C}{4} fit are duplicated for \ion{Si}{4} but with column densities $N_{\rm SiIV}$ adjusted to match the optical depth of the detected \ion{Si}{4} components. The redshifts of the various sub-components in \ion{C}{4} and \ion{Si}{4} are tied, and the parameters of the initial guesses are again passed to VPFIT. VPFIT then solves for the tied redshift of each component as well the column densities ($N_{\rm CIV}$ and $N_{\rm SiIV}$) and Doppler widths ($b_{\rm CIV}$ and $b_{\rm SiIV}$) which are allowed to vary independently. Solutions with matching subcomponent structure in both ions are preferred; however, in some cases absorption is detected in one ion only. For such absorbers, the undetected component is dropped from the fit to the ion in which it is not detected. 

Absorbers fit jointly in \ion{Si}{4} and \ion{C}{4} typically have Doppler widths with $b_{\rm CIV} \gtrsim b_{\rm SiIV}$, as expected in the case of thermal broadening because C is lighter than Si. As \ion{Si}{4} absorbers are often found to be narrower than their \ion{C}{4} counterparts, occasionally, unrecognized subcomponents are identified in the \ion{Si}{4} data, in which case the component is added to both the \ion{Si}{4} and \ion{C}{4} fits and VPFIT is rerun. 

Once a satisfactory fit is achieved for \ion{Si}{4} and \ion{C}{4}, their Doppler widths are used to determine the temperature ($T$) and turbulent velocity ($v_{\rm turb}$) of each component using equations \ref{temp_turb1} and \ref{temp_turb2}.
The Doppler widths of the two transitions are then tied with a single value of the temperature ($T$) and turbulent velocity ($v_{\rm turb}$) for each component and VPFIT is rerun providing measurements and uncertainties for the the physical parameters ($T$ and $v_{\rm turb}$) rather than the model parameters ($b_{\rm CIV}$ and $b_{\rm SiIV}$). 

Using the output of this round of fitting, duplicate components are added in lower-ionization stages where detected (typically some subset of \ion{C}{2}, \ion{C}{3}, \ion{Si}{2}, and \ion{Si}{3}) now with tied redshifts, $T$, and $v_{\rm turb}$.\footnote{Note that all ionization states detected in a given component are fit assuming the same value of $T$ and $v_{\rm turb}$.} The column densities of the newly added ions are again scaled to roughly match the observed optical depths in those species, and the full set of estimates are again run through VPFIT. As with the combined \ion{Si}{4} and \ion{C}{4}, occasionally components need to be added in order to produce a satisfactory fit to the lower-ionization data and undetected components in individual ions are dropped from the fit to that ion. 

In one instance, (Q0100-BX210 which has a particularly complex sub-component structure) VPFIT would not converge when allowed to fit for $T$ and $v_{\rm turb}$; however, in this system, the determined $b_{\rm CIV}/b_{\rm SiIV} \approx \sqrt{m_{\rm Si}/m_{\rm C}}$, the ratio expected in the case of pure thermal broadening, so the fit for that system was run assuming  $b_{\rm CIV}/b_{\rm SiIV} = \sqrt{m_{\rm Si}/m_{\rm C}}$. This represents 8/130 of the absorbers reported in this work. Similarly, in 7 absorbers with weakly detected and highly blended metal lines, the determined $b_{\rm CIV}$ and $b_{\rm SiIV}$ were roughly equal and were fixed to purely turbulent broadening to improve the convergence of VPFIT.

We favor Voigt profile fits that have the minimum number of sub-components required to produce an adequate fit. However, given the complex nature of some of the absorption systems seen in the CGM, unrecognized blended components are possible. Given that a large fraction of the absorbers with multi-element constraints are consistent with nearly pure thermal broadening (see Section \ref{temp}), we expect that the effect of unrecognized blends is minimal. 

The final result of the fitting process is a list of the detected column densities of the various ionic transitions for each subcomponent as well as their redshift, gas temperature $T$, and turbulent velocity $v_{\rm turb}$. In cases where a sub-component is detected for only one element, $T$ and $v_{\rm turb}$ are not constrained, and only $N$, $z$, and $b_{\rm d}$ are measured.

\subsection{\ion{O}{6}}
\label{O6}

The final step of the Voigt profile decomposition deals with \ion{O}{6}, the highest ionization potential species detected in these data. \ion{O}{6} is a challenging ion to constrain in the high-$z$ CGM because the wavelength of the \ion{O}{6} doublet $\lambda\lambda 1031,1037$\AA\ is very similar to the wavelength of the Lyman $\beta$ transition ($\lambda 1025$\AA) which is ubiquitous in the high-$z$ IGM. Thus, in many cases, our data are too contaminated by high-column density \ion{H}{1} systems to measure \ion{O}{6}; however, the doublet is detectable and quite strong in others. As shown in Figure \ref{BX182}, the spectral morphology of \ion{O}{6}, where detected, is very different from that seen in the ions with lower ionization potential. In particular, while the lower ions show many narrow subcomponents, the \ion{O}{6} section of the spectrum typically exhibits a handful of very broad absorbers which are less apparent in the lower ions.

However, in the systems with the strongest detected  \ion{O}{6} components, the \ion{C}{4} fit described in  the previous section (which has no constraint provided by the \ion{O}{6} portion of the spectrum) typically contains several broad components that are well matched with those seen in \ion{O}{6} but with significantly lower optical depths (see Figure \ref{BX182}). Given that the \ion{O}{6} region of the spectrum is not included as a constraint in the \ion{C}{4} fit up to this point, the data in the \ion{C}{4} region independently prefer the presence of broad components. 

Given that the broad \ion{C}{4} components typically overlap in redshift space with the narrower components, their fits are often uncertain. In order to better constrain the properties of these broad absorbers, we introduce the \ion{O}{6} section of the spectrum to the fit and tie the redshifts and Doppler parameters of broad \ion{O}{6} components to broad lower ionization lines where detected and rerun VPFIT.

\begin{deluxetable}{lrrr}
\tablecaption{Number of Components in Voigt Profile Decomposition}  
\tablewidth{0pt}
\tablehead{
\colhead{Ion} & \colhead{Median\tablenotemark{a}} & \colhead{Mean\tablenotemark{a}} & \colhead{Max}}
\startdata
 \ion{Si}{2} &           13 &           15 &           20 \\
\ion{C}{2} &           14 &           15 &           18 \\
\ion{Si}{3} &            8 &            9 &           10 \\
\ion{C}{3} &           11 &            9 &           14 \\
\ion{Si}{4} &           12 &           10 &           20 \\
\ion{C}{4} &           15 &           14 &           22 
\enddata
\tablenotetext{a}{The median and mean refer to the number of components per galaxy among those galaxies in which the ion was detected. Note that for each ionization state, the sample of galaxies in which gas is detected is different (see Table \ref{table_N}).}
\label{components}
\end{deluxetable}

\begin{figure*}
\center
\includegraphics[width=0.5\textwidth]{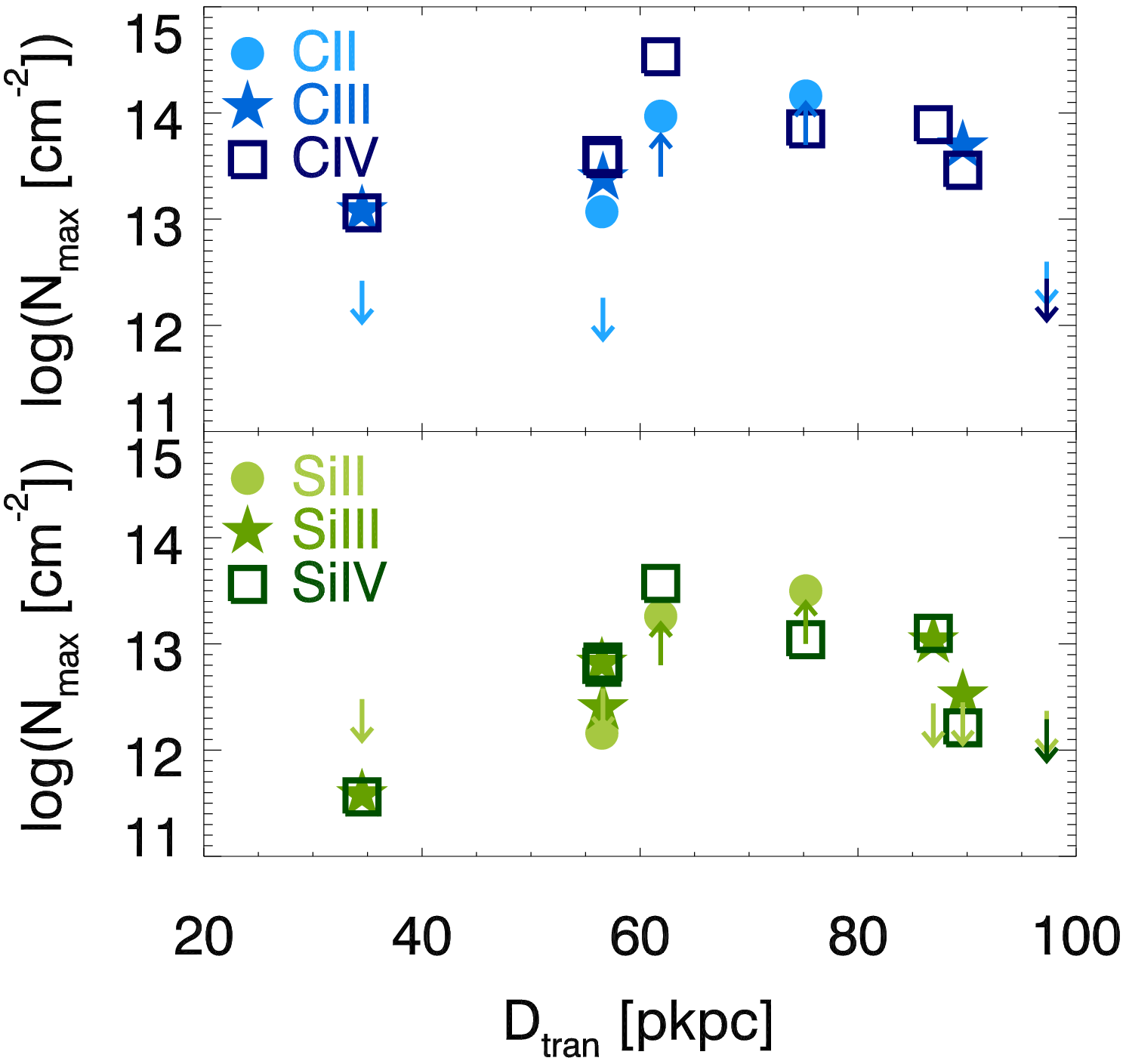}\includegraphics[width=0.5\textwidth]{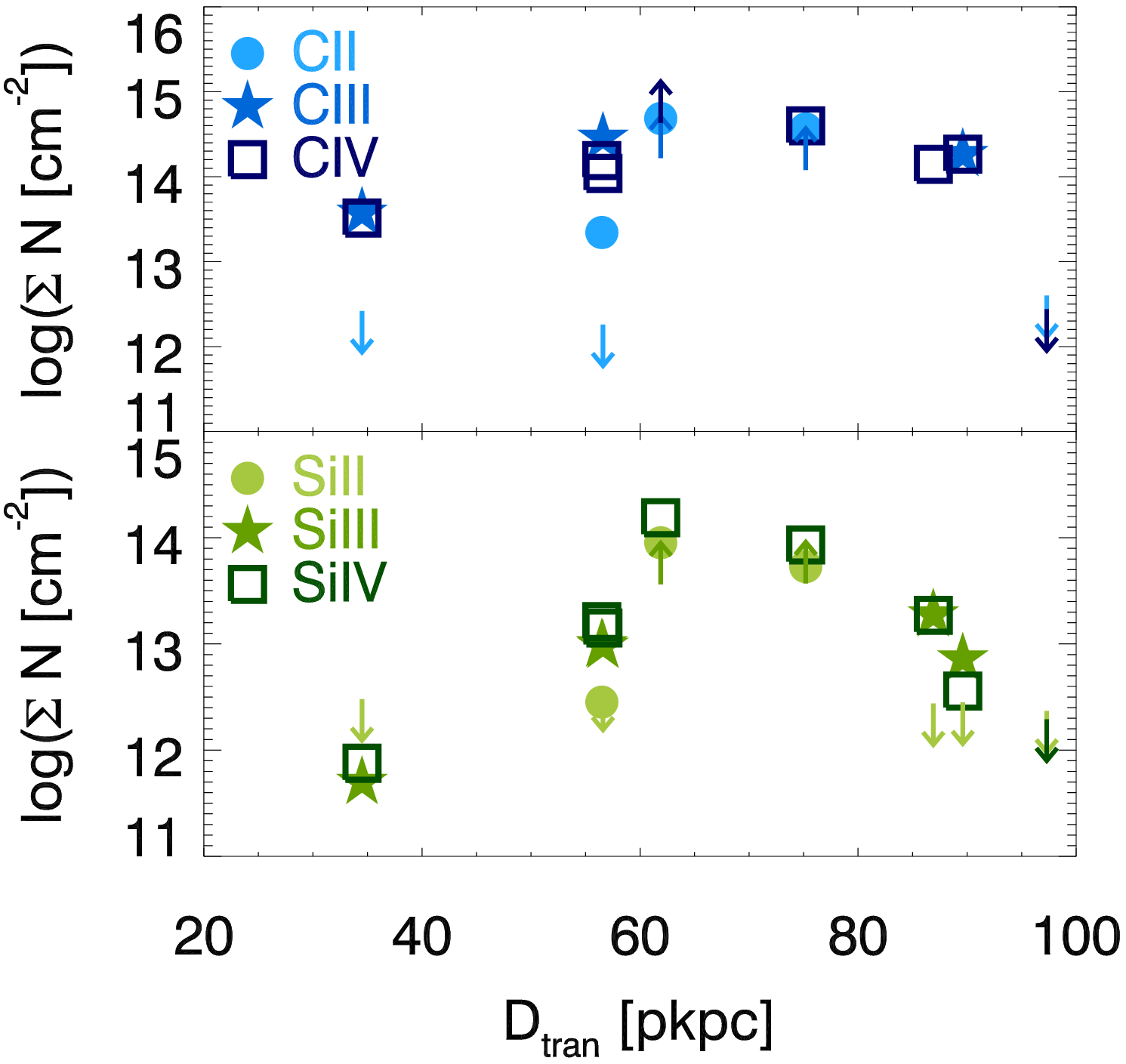}
\caption{The highest column density absorber, max($N_x$) [left panel] and the total column density, $\Sigma{N_x}$ [right panel] in each of 6 species as a function of impact parameter between the galaxy and the line of sight to the QSO. Downward pointing arrows correspond to non-detections while upward-pointing arrows represent saturated absorbers with lower-limits on the column density. Note that due to the similar impact parameters of Q1442-MD50a and Q1549-D15, the column densities corresponding to those galaxies appear overplotted. The closest galaxy in the sample, Q0105-BX90b, is likely lower-mass than the rest of the KBSS galaxies (see text for details), and so its relatively low column densities may not be typical of the population. All galaxies within 90 pkpc have absorption in one or more ionization state of C and Si. Halo gas at $z\sim2$ contains large column densities of metal ions.}
\label{maxNsumN}
\end{figure*}

There are no visually apparent broad components in ions with ionization potentials below that of \ion{Si}{4}. Further, the expectation (given their very different ionization potentials) would be that \ion{O}{6} bearing gas would not contain singly ionized species. However, given the fitting procedure defined above, in some cases the fit to low ionization species such as \ion{C}{2} and \ion{Si}{2} also includes broad components while in others these components have been dropped by VPFIT. 

We use non-equilibrium cooling models to determine the expected ionic phases for broad components consistent with high gas temperatures. For initially hot gas radiatively cooling to $T\lesssim10^6$K, non-equilibrium ionization effects are likely due to the short cooling times at these temperatures. Rapid radiative cooling leads to the recombination of the gas lagging behind the drop in temperature resulting in over-ionization of the gas compared to collisional ionization equilibrium (CIE) predictions (see e.g. \citealt{gna07,opp13}). \citet{opp13} and \citet{gna17} revisited this non-CIE calculation including the presence of radiation. 

Based on these comparisons to the models from \citet{gna17}, we remove all of the low-ionization fit components with inferred temperatures $\log(T/ \rm K) \gg 5$. For absorption components with $\log(T/ \rm K) \approx 5$ and those with $\log(T/ \rm K) \gg 5$ detected in the doubly ionized phase and/or \ion{Si}{4}, we consider the column density ratios of various ions and their errors as well as the error on the temperature measurements. In cases where the inferred ion ratios and temperatures are implausible for all model tracks, we again remove that component of the fit to the lower-ionization species in question. Once components with unphysical parameters are removed, we re-run VPFIT. The output of this final run forms the basis for the conclusions in this paper.

\section{Column Densities and Covering Fractions}

\label{cf_section}

The gas within the CGM of high-$z$ galaxies is kinematically complex and multi-phase (see Figure \ref{BX182}). We commonly detect gas from singly ionized through triply ionized species at the same velocities and with overall similar profiles. Higher ionization gas, such as that exhibiting \ion{O}{6} absorption, is less kinematically complex and bears little resemblance to the lower-ionization gas, implying a very different origin.

The Voigt profile decomposition of metal line detections in the sample vary in their complexity. Individual galaxies in the sample contain anywhere from 0 to 31 metal line components with a median (mean) of 15 (19) components per galaxy (Table \ref{table_N}). Broken down by ionization state, galaxies with detected absorption in a given ion typically have $>10$ components within $\pm1000 $ \kms\ (Table \ref{components}).

Recently \citet{pee18} created mock spectra of sightlines passing through a zoom-in simulation of a Milky Way-like halo at $z>2$ in the FOGGIE simulation. FOGGIE is one of several recent simulations employing enhanced halo resolution (forced refinement within the CGM, \citealt{hum18}), allowing them to resolve the detailed structure of halo gas \citep{hum18,pee18,van18,sur19}. \citet{pee18} do not report the general statistics for the number of components found along randomly drawn sightlines, instead focusing on the number found along simulated sightlines containing high column density \ion{H}{1} absorbers ($N_{\rm HI}>10^{17}$ \cm2, Lyman Limit Systems, LLS). In these simulations, the median number of \ion{C}{4} (\ion{Si}{4}) components found along LLS-bearing sightlines is 3 (5), significantly lower than our median number of components of 15 (12). This comparison suggests there may be CGM structures that remain unresolved even in simulations such as FOGGIE \citep{hum18}; however, given the mismatch in the selection of lines of sight in FOGGIE and the sample presented here, a more direct comparison should be made before stronger conclusions can be drawn.

\subsection{Column Densities}

To quantitatively assess the frequency of detection of multiple species and the quantity of gas found in each ionization state, here we present statistics based on both the maximum column density absorber in each ion, max($N_x$), and the total column density in each ion, $\Sigma{N_x}$ (Table \ref{table_N}). The column densities range from non-detections to $\log{(\Sigma{N_x})}>14.6$. For one galaxy (Q1623-BX432), we detect no metal absorption down to relatively sensitive limits ($\log(N_x)\lesssim 12.5$, see Figure \ref{maxNsumN} and Table \ref{table_N}). While this is the galaxy with the largest impact parameter presented here, we note that strong metal line absorption is detected within the more-extended halos of many galaxies at larger impact parameters within the KBSS (Rudie et al. in prep). 

Figure \ref{maxNsumN} shows the statistics for max($N_x$) and $\Sigma{N_x}$ as a function of impact parameter. KBSS galaxies have large quantities of ionized metals in their CGM, with a typical $N_{\rm CIV} > 10^{14}$ \cm2. Singly ionized species are more commonly detected closer to galaxies. We do not detect singly ionized C or Si beyond 75 pkpc in this sample. However, all galaxies within 90 pkpc have detected absorption in one or more ionization state of C and Si.

With $D_{\rm tran} =$35 pkpc, Q0105-BX90b has the smallest impact parameter in the current sample; however, as discussed in Section \ref{observations}, it also has a significantly lower stellar mass than the other sample galaxies. Therefore, the low measured column densities of metallic ions at small galactocentric distance are probably not representative of more typical KBSS galaxies. However, even if Q0105-BX90b is removed, the maximum and total column densities are not monotonic or strongly correlated with impact parameter within $R_{\rm vir}$. This is unsurprising given the small sample size and small dynamic range in impact parameter presented here. Future work will present measurements of \ion{C}{4} and \ion{Si}{4} at larger distances which do show the expected strong anti-correlation between impact parameter and column density, although with considerable scatter (Rudie et al, in prep). 

For well-detected and non-saturated absorbers, max($N_x$) and $\Sigma{N_x}$ are well defined. However, in the case of saturation or non-detections, we place lower or upper limits, respectively, on max($N_x$) and $\Sigma{N_x}$ as described in the next section.

\begin{figure*}
\center
\includegraphics[width=0.8\textwidth]{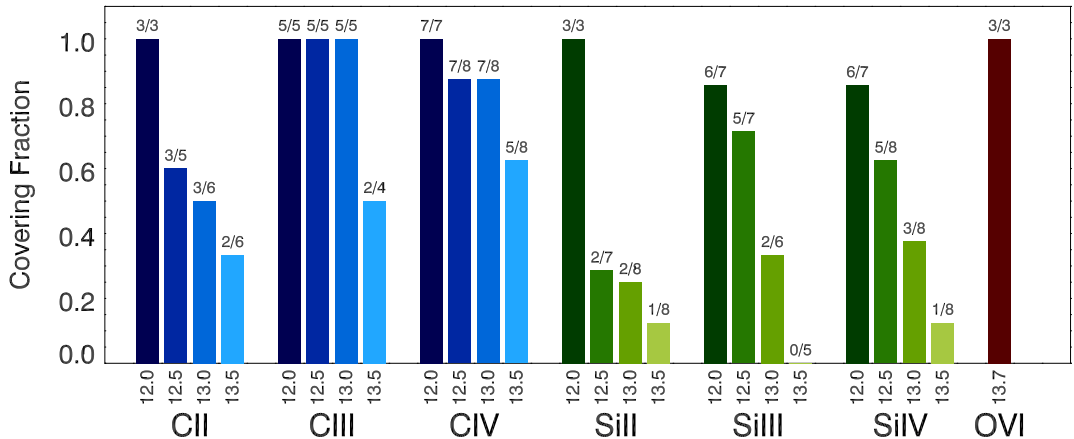}
\caption{Covering fractions of various ionic species within 100 pkpc and $\pm 1000$ \kms of galaxies in the KBSS. These values are determined using the column densities of individual components in the Voigt profile decomposition. Note that in systems with the most sensitive data, all the ionization states of carbon are always detected. For higher column densities [$\log(N_x/ {\rm cm}^{-2}) = 12-12.5$] at which all of the uncontaminated sightlines in the sample are sensitive, the covering fraction for all three ionization stages of carbon is $>50\%.$ \ion{Si}{3} and \ion{Si}{4} show equally high covering fraction, only \ion{Si}{2} is detected less frequently than 50\% within the halo of these galaxies.}
\label{cf}
\end{figure*}

\begin{figure*}
\center
\includegraphics[width=0.8\textwidth]{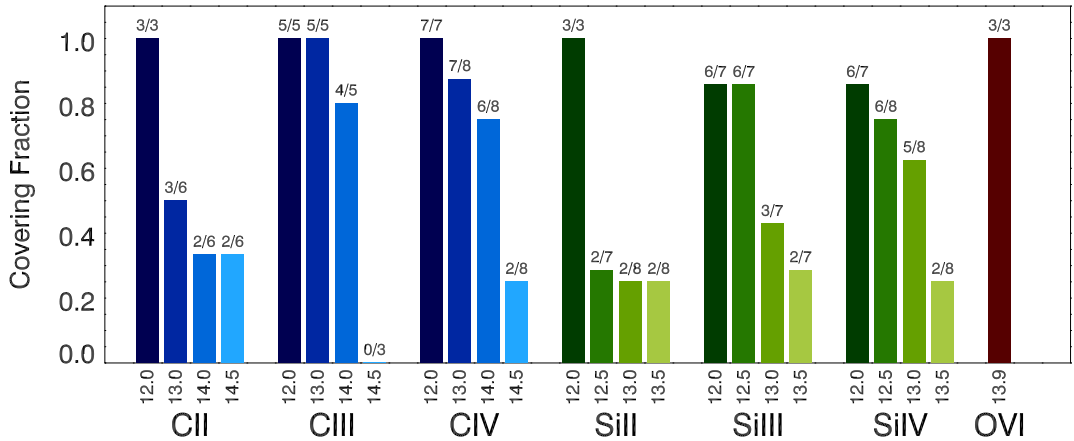}
\caption{Same as Figure \ref{cf} above (the covering fraction of various species within 100 pkpc), but calculated for the total column density within $\pm 1000$ \kms, $\Sigma N_{\rm ion}$.}
\label{cf_sumN}
\end{figure*}

\subsection{Non-detections and Saturated systems}

\label{non-det}

For ions in which all observed transitions are saturated (or for singlets that are saturated), we estimate a lower limit on the total column density in the saturated velocity range by computing the equivalent width in the saturated regions and converting it to a column density using the linear portion of the curve of growth. This lower limit on the column density is treated as a single absorber and added to the catalog. To compute $\Sigma{N_x}$, we add the determined $N_x$ to that of all of the unsaturated components, and report the total column density as a limit. 

For ions where some or all of the transitions are uncontaminated, but where the ions are still not detected, we compute upper limits on the column density. These limits are derived by computing the limiting equivalent width within the velocity range surrounding each galaxy in which the transmission in the Ly$\alpha$ regions drops below 80\%. For highly contaminated regions, we do not report a column density limit.

\subsection{Covering Fractions}

With column density measurements and limits in hand, we can also calculate the covering fraction of halo gas as a function of ionization state and limiting column density. Figures \ref{cf} and \ref{cf_sumN} show the measured covering fraction of gas of various ionization states within 100 kpc of galaxies in the KBSS. 

The covering fraction of gas is defined as the number of galaxies with detected absorption above a limiting column density divided by the number of galaxies in which it would be possible to detect absorption above that limit. In Figures \ref{cf} and \ref{cf_sumN} the value of the number of galaxies with detections and the number of galaxies where the measurement is possible is printed above the bar in each bin of column density and ion. 

To measure the covering fraction, we use the upper limits derived for non-detections in Section \ref{non-det} to determine if the absorption line data for a given galaxy is sufficient to detect absorbers of a given $N_x$. Galaxies without detections that (a) have spectral coverage of the ion in question, (b) are not strongly contaminated and (c) have sufficiently sensitive spectra contribute to the denominator of the covering fraction for a given limiting column density (the x-axis in Figures \ref{cf} and \ref{cf_sumN}). For this reason, as the value of the limiting $N_x$ declines, the number of galaxies with sufficiently sensitive spectra decreases as well, and so the value of the denominator declines. 

Note that for some ionization states, there are galaxies without spectral coverage of some transition (\ion{C}{3}). Similarly, for some galaxies, the absorption line spectra of particular ions (\ion{C}{2}, \ion{C}{3}, \ion{Si}{3}, \ion{O}{6}) are too heavily contaminated to be useful probes of the covering fraction. These galaxies do not contribute to the measurement of the covering fraction for any ions in which their absorption-line spectra are insensitive. 

Figure \ref{cf} shows the covering fraction as a function of the column densities of individual absorbers while Figure \ref{cf_sumN} considers the same statistic for $\Sigma{N_x}$ within $\pm1000$ \kms. Note that at low column densities [$\log(N_x/ {\rm cm}^{-2}) = 12-12.5$], the covering fraction for doubly and triply ionized species is very high, $87-100\%$. At higher column densities, the covering fractions decline, however, the covering fraction of \ion{C}{4} remains $>50\%$ even for $N_{\rm CIV} \ge 10^{13.5}$ \cm2.

\subsection{OVI}

\label{cf_OVI}

Because \ion{O}{6} is located within the Ly$\beta$ forest, it suffers significantly more contamination than other ions discussed in this work. In velocity ranges where \ion{H}{1} and \ion{C}{4} absorption are detected and where the \ion{O}{6} portion of the spectrum is not highly contaminated (3 galaxies), there are clear detections of \ion{O}{6} with relatively high $N_{\rm OVI}\sim 10^{14}$ \cm2. In velocity ranges with greater contamination, all of the sightlines are consistent with having \ion{O}{6} absorption with $N_{\rm OVI} > 10^{14}$ \cm2; however, in such cases, the component structure of \ion{O}{6} absorption, if present, is not sufficiently constrained for fitting. The data are therefore consistent with a 100\% covering fraction for $N_{\rm OVI}\ge 10^{13.7}$ \cm2 and $\Sigma N_{\rm OVI} \ge 10^{13.9}$ \cm2 as shown in Figures \ref{cf} and \ref{cf_sumN}. 

While the data are often sufficiently contaminated that it is challenging to make unambiguous measurements of the \ion{O}{6} column, there are reasons to suspect that there is \ion{O}{6} absorption and that the column densities may typically be large ($N_{\rm OVI}\sim 10^{14}$ \cm2). First, we compute limits on the \ion{O}{6} column density by measuring the equivalent width of possible \ion{O}{6} in the velocity regions with high \ion{H}{1} opacity within $\pm1400$ \kms\ of the systemic velocity of the galaxies. In 7/8 of the galaxies, the velocity regions containing the highest inferred \ion{O}{6} equivalent widths were the velocity regions closest to the systemic redshift of the galaxy. If all of the absorption in the \ion{O}{6} regions were due to contamination, one would expect more scatter in the velocity range of the highest apparent opacity. 

Other circumstantial evidence of \ion{O}{6} absorption comes from a comparison of the \ion{O}{6} section of the spectrum with that of \ion{C}{4}. If one assumes \ion{O}{6} is coincident in velocity space with \ion{C}{4} absorption, as is the case for regions where \ion{O}{6} is detected with high confidence, the data are typically consistent with $N_{\rm OVI}/N_{\rm CIV} \geq 1$. 

Additional support for the common existence of a substantial \ion{O}{6} bearing phase comes from a pixel optical depth analysis of the KBSS sample. \citet{tur15} found strongly enhanced \ion{O}{6} absorption at fixed \ion{H}{1}, \ion{C}{4}, and \ion{Si}{4} optical depths within 180 pkpc and $\pm$350 \kms\ of KBSS galaxies compared to random locations in the spectra. The strength of the detected \ion{O}{6} absorption and the high \ion{O}{6} to \ion{H}{1} optical depth ratios found imply that the most plausible excitation mechanism for some of the \ion{O}{6}-bearing gas is collisional ionization in a $T>10^5$ K gas with metallicity of at least 10\% solar. Given the high inferred metallicity, velocities in excess of the circular velocity of the halo, and that these conditions were found uniquely in regions close to galaxies,  \citet{tur15} conclude the most likely source would be galactic wind material. The \ion{O}{6} absorption shown in Figure \ref{BX182} and clearly detected in the other uncontaminated systems is consistent with this interpretation.

\subsection{Comparison to theory and other observations}

\citet{she13} computed the covering fraction of various ions surrounding the Eris2 simulated galaxy. Eris2 is matched to the Milky Way mass at $z=0$ and so is certainly less massive than the majority of the KBSS galaxies. \citet{she13} report the covering fractions for $N_x>10^{13}$ \cm2\ for each ion within $R_{\rm vir}$. Compared with our observations (Figure \ref{cf_sumN}), they find a consistent covering fraction of \ion{Si}{4} ($C_f\sim70\%$), \ion{C}{4} ($C_f\sim90\%$), and \ion{O}{6} ($C_f=100\%$), but their simulations over-produce the covering fraction of \ion{C}{2} ($C_f\sim70\%$) and \ion{Si}{2} ($C_f\sim40\%$) compared to the KBSS observations ($C_{f, \rm CII}=50\%$ and $C_{f, \rm SiII}=30\%$) within $R_{\rm vir}$. 

\citet{lia16} analyzed simulations which evolved into a Milky Way mass galaxy at $z=0$ (lower mass than is typical of the KBSS at $z\sim2.3$), considering the radial profile of \ion{H}{1} and metals surrounding galaxies as a function of redshift. For the snapshots at $z=2$ and $z=3$ compared at radial distances scaled by the virial radius of the halo, the typical column densities seen in the simulations underpredict that seen in the KBSS data for \ion{H}{1} and \ion{C}{4}. For \ion{H}{1} \citet{gcr12} report a typical $N_{\rm HI}=10^{16.5}$~\cm2 within $0.5<R/R_{\rm vir}<1.0$ where as the simulations of \citet{lia16} have typical $N_{\rm HI}<10^{14}$~\cm2 at $z=2$ and $10^{14}<N_{\rm HI}<10^{15}$~\cm2 at $z=3$. For \ion{C}{4}, over half of the KBSS sightlines within $R_{\rm vir}$ have $N_{\rm CIV}>10^{14.0}$~\cm2 while \citet{lia16} find $N_{\rm CIV}<10^{12.0}$~\cm2 at $z=2$ and $N_{\rm CIV}<10^{13.0}$~\cm2 at $z=3$ for all of their simulations except for one. At $z=3$, the \ion{C}{4} column densities predicted for a model with more energetic supernovae are somewhat higher ($10^{12}<N_{\rm CIV}<10^{14.5}$~\cm2), closer to the observed value. Similarly, the \ion{O}{6} column densities are somewhat lower than observed in all models ($10^{13}<N_{\rm OVI}<10^{14}$~\cm2 at $z=2$ and $z=3$ whereas KBSS galaxies have $N_{\rm OVI}>10^{13.9}$~\cm2), with the closest model being the $z=3$ snapshot with more energetic SNe ($10^{13.5}<N_{\rm OVI}<10^{14}$~\cm2). 

Again we note that the mass scale of the \citet{lia16} and \citet{she13} simulations were meant to match a Milky Way progenitor and are therefore less massive than is typical of the KBSS galaxies. This limits the utility of detailed comparisons with these models. Simulations with galaxy properties more similar to the KBSS would be more useful in diagnosing any discrepancies between the data and current models which could lead to better understanding of the nature of accretion and feedback in the distant universe. 

The measured \ion{O}{6} column densities ($N_{\rm OVI}>10^{13.9}$~\cm2) are similar to the typical \ion{O}{6} columns found surrounding low-redshift galaxies \citep{wer13,joh15,joh17,zah18} and also the typical column densities of low-redshift \ion{O}{6} systems found in random sightline surveys \citep{tho08,che09,sav14}. Several theoretical models have successfully reproduced these high column densities, using a variety of physical scenarios and assumptions \citep{hec02,she13,fae17,bor17,opp18,voi18,qu18,mcq18,ste18}, suggesting the column densities alone may not be sufficient to discriminate between models and that more detailed observations which further constrain the nature of \ion{O}{6} absorbers are needed. In Section \ref{temp}, we discuss constraints on the widths of \ion{O}{6} absorbers, and temperatures where possible, in a first attempt to provide these more-detailed constraints.

\section{Gas Kinematics}

\label{kinematics}

\begin{figure}
\center
\includegraphics[width=0.45\textwidth]{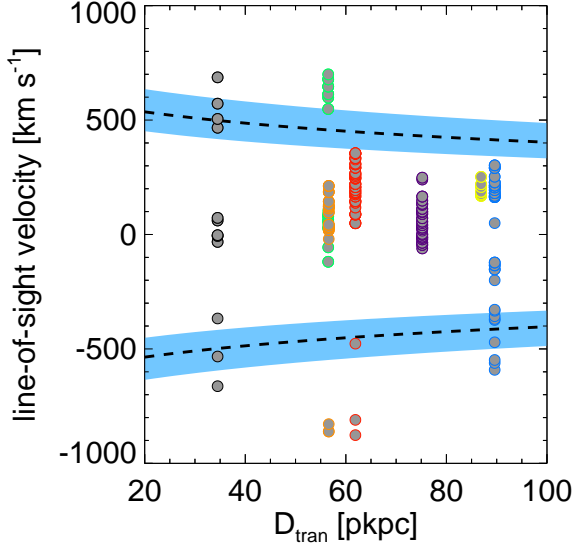}
\caption{The velocities and impact parameters of metal-enriched absorbers within 100 pkpc of a galaxy in the KBSS sample. Errors in velocity and impact parameter are smaller than the symbol size. Each distinct galaxy is plotted with a different color outer annulus so that the points corresponding to Q1442-MD50a and Q1549-D15 can be differentiated. Plotted for comparison is the 3D escape velocity as a function of distance for halos of $\log({M_{\rm halo}/M_{\odot}})=11.9 $ (black dashed line) and $11.7<\log({M_{\rm halo}/M_{\odot}})<12.1$ (blue shading). 
Notably, for 5/7 galaxies, we detect metal enriched absorbers at velocities well in excess of the escape velocity implying that some of the detected CGM is unbound to the galaxies.}
\label{vesc}
\end{figure}

The kinematics of gas in the CGM of high-$z$ galaxies offers important clues about the baryon cycle. One quantity of particular importance is the fraction of mass and metals that are ejected from the ISM of the galaxy through galactic winds, many of which reside within the CGM. Additionally, understanding the fate of galactic winds once outside the ISM is critical; the chemical evolution of galaxies is strongly coupled to the amount of metals lost from the system via galactic winds versus the amount that may return to the galaxy as `recycled' material (see e.g. \citealt{sha76}; \citealt{opp10}; \citealt{ang17}).

Assessing the fate of metal enriched material is complicated by our limited knowledge of the 3D kinematics of gas with respect to galaxies. In particular, in CGM studies, we can only observe the redshift of absorbers from which we derive constraints on both the radial velocity of the gas and its position along the line of sight. Due to the ambiguity between position and velocity, it is not generally possible to determine the \textit{direction} of motion of an individual absorber into or out of the galaxy. Past work has instead relied on ensemble statistics and comparisons to analytic scalings such as the circular velocity ( e.g., \citealt{gcr12,rak12}.)

Given that we can only measure the \textit{line-of-sight velocity} of the gas, the measured difference in redshift between the galaxy and the absorber, $\Delta v$,  provides only a lower limit on the true 3D velocity of the gas. The only unambiguous interpretation is for measured line-of-sight velocities larger than the escape velocity at a given projected distance ($\Delta v > v_{\rm esc}$). In such systems, we can say with certainty that the gas is not bound to the galaxy, although it is still not possible to know with certainty that the gas originated within the galaxy in question.

Figure \ref{vesc} shows the measured velocities of absorbers with respect to the systemic velocity of their closest known galaxy, and how the velocities relate to the impact parameter between the galaxy and the absorbers. When interpreting this figure, it is important to keep in mind that the impact parameter, $D_{\rm tran}$ is always a lower limit on the 3D distance between the galaxy and the absorber, and the velocity is always also a lower limit on the true 3D velocity of the gas with respect to the galaxy. 

\begin{figure}
\center
\includegraphics[width=0.45\textwidth]{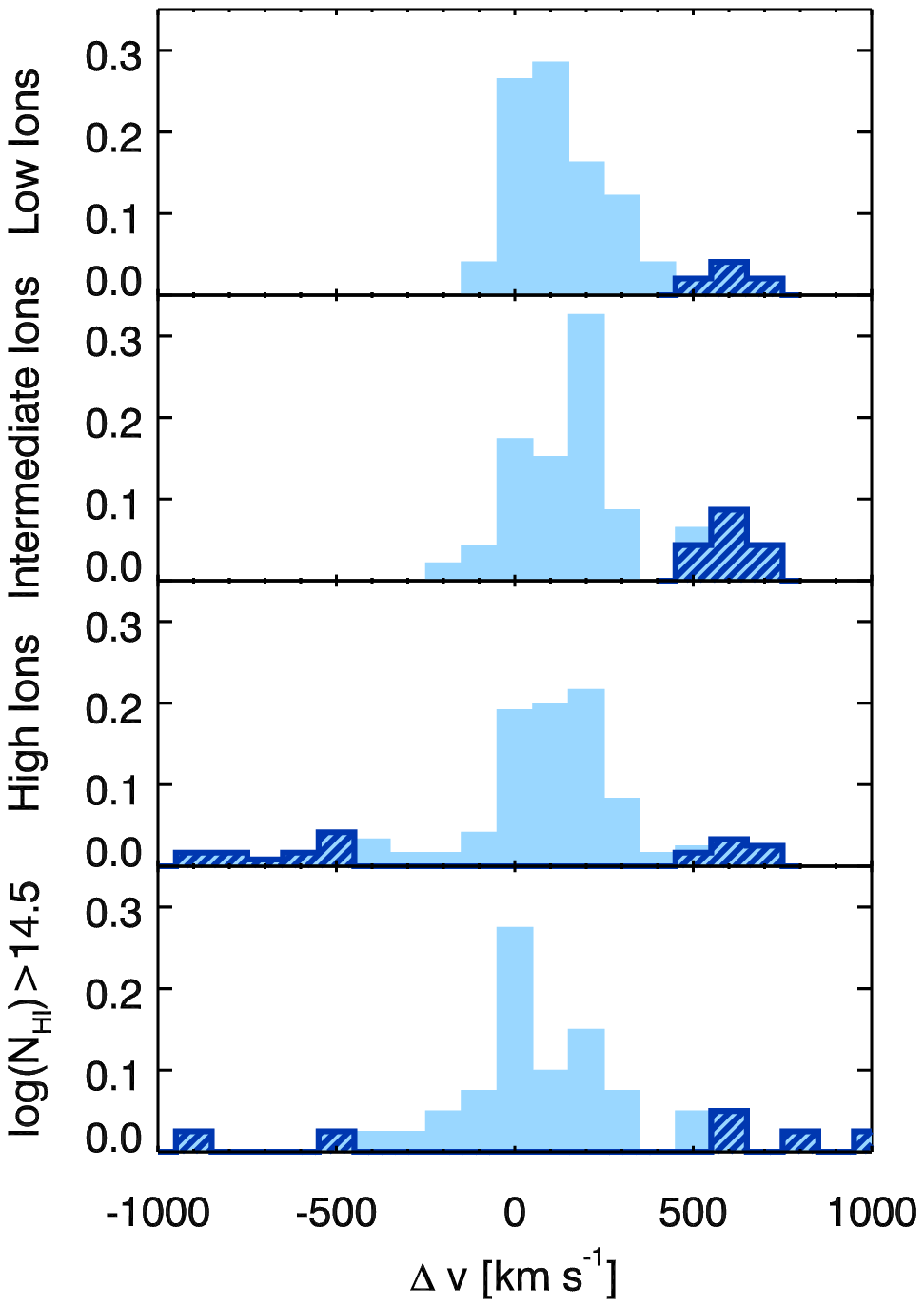}
\caption{The normalized velocity distribution of absorbers of various ionization potentials. Low ions refers to absorbers with detected \ion{C}{2} or \ion{Si}{2} absorption, intermediate ions are \ion{C}{3} or \ion{Si}{3}, and high ions are \ion{C}{4}, \ion{Si}{4}, \ion{N}{5}, or \ion{O}{6}. Also included for comparison are high-$N_{\rm HI}$ absorbers [with $\log(N_{\rm HI}/{\rm cm}^{-2}) > 14.5$] which were shown in \citet{gcr12} to correlate strongly with the positions of galaxies. The dark blue hatched histogram represents those absorbers with $\Delta v> v_{\rm esc}$ for a dark matter halo mass $\log(M_{\rm halo}/M_{\odot})=11.9$ as shown in Figure \ref{vesc}.  Note that the velocity distribution of high ions is more extended towards higher velocities, and that a higher fraction of the unbound absorbers are detected in high ions. Most of the unbound absorbers do not show strong \ion{H}{1} absorption as can be seen in the bottom panel. The bias towards redshifted systems relative to the galaxies' systemic velocities is likely due to the small number of galaxies (7) contributing to this measurement.}
\label{kinematic_hist}
\end{figure}

We compare the measured velocities of absorbers to the expected escape velocities of halos in the following way. We compute the escape velocity as a function of 3D distance from an NFW halo \citep{nfw} with virial masses of $11.7<\log{(M_{\rm halo}/M_{\odot})}<12.1$. We assume a halo concentration of 3.0 following \citet{duf08}. The mass range is selected to be the range of halo masses that reproduces the measured clustering of KBSS galaxies \citep{tra12}. \footnote{\citet{tra12} report that halos with a minimum halo mass M$_{\rm halo, min}=10^{11.7}$ M$_\odot$ reproduce the clustering of the KBSS galaxies. Based on the halo mass function above M$_{\rm halo, min}=10^{11.7}$ M$_\odot$, the median mass is M$_{\rm halo}=10^{11.9}$ M$_\odot$, and we choose $11.7<\log{(M_{\rm halo}/M_{\odot})}<12.1$ to encompass the majority of halos of galaxies in the sample. \citet{rak13} used \ion{H}{1} absorption in the CGM in comparison with hydrodynamic simulations, finding M$_{\rm halo, min}=10^{11.6}$ M$_\odot$, consistent with \citet{tra12}. If we instead use the stellar mass-halo mass relation at $z=2$ from \citet{beh13}, and the maximum mass of the galaxies in the sample M$_{*}=10^{10.7}$ M$_\odot$ we would compute a maximum halo mass of M$_{\rm halo}=10^{12.2}$ M$_\odot$ which would not change the results presented herein.} The escape velocities are plotted in Figure \ref{vesc} as the blue band. The black dashed curve highlights the escape velocity for a $\log{(M_{\rm halo}/M_{\odot})}=11.9$ halo, the median halo mass of galaxies in our sample. 

Notably, 5/7 (70\%) of the galaxies for which we detect metals within the virial radius exhibit metal enriched gas at velocities well in excess of the escape velocity of their halos. For points that lie outside the blue shaded regions, we can say \textit{unambiguously} that the gas is not bound to the galaxy in question. This is in contrast to low-redshift observations which have generally found absorbers with measured radial velocities well below the escape velocity of their halos \citep{tum11,sto13, zhu14,san16,hua16,wer16} although see \citet{tri11} and \citet{san16} for some examples of gas with $\Delta v > v_{\rm esc}$.

As with all CGM metal-bearing absorption systems, it is possible that the metals in the unbound absorbers were not ejected from the galaxies in question, but arise from some other source.\footnote{We note that within 100 kpc, we have spectroscopic redshifts for the majority of galaxies that meet our photometric selection criteria. As such, if these high-velocity absorbers are due to other clustered galaxies, they would have to be due to lower-luminosity galaxies than those considered here.} But given the high fraction of KBSS galaxies with detected unbound absorbers, we consider the probability of such a scenario to be low. 

Further, the velocities of the unbound absorbers are not unexpected. KBSS galaxies commonly show strong signatures of outflowing ISM, detected through blue-shifted UV interstellar absorption lines \citep{sha03,ccs10}. KBSS rest-UV spectra typically show outflow velocities as high as $800$ \kms\ in low-ionization gas as well as high-ionization species such as \ion{C}{4}. The location of this absorbing gas is uncertain, but it is certainly plausible that the unbound gas we detect within the halo results from the high-velocity outflows commonly seen in these galaxies. 

While it is not possible to say definitively which absorbers are bound to the galaxy\footnote{Absorbers in Figure \ref{vesc} that lie between the two curves have more ambiguous kinematics. Because $D_{\rm tran}$ is a lower-limit on the 3D distance and $\Delta v$ is a lower limit on the 3D velocity of the absorbers, those absorbers who appear to have $|v|<v_{\rm esc}$ could truly be bound to the galaxy, or they could be more distant and/or moving sufficiently fast along the plane of the sky that they are actually unbound.}, the nature of the unbound absorbers is of considerable interest. 
In Figure \ref{kinematic_hist} we compare the velocity distribution of absorbers of a variety of ionization stages to understand if absorbers at high velocity are different from those closer to the systemic velocity of the galaxy. The top panel show singly ionized species, followed by doubly ionizes species, while the two lower panels show highly ionized metal species (\ion{Si}{4}, \ion{C}{4}, \ion{N}{5}, \ion{O}{6}) and absorbers with $\log(N_{\rm HI}/{\rm cm}^{-2})>14.5$. Notably,  the velocity distribution of highly ionized absorbers extends to higher overall velocities; however, the general velocity distribution of absorbers is similar. For low, intermediate, and highly ionized absorbers, $8\pm 4\%, 17\pm6\%$, and $18\pm 4\%$ of absorbers have $|\Delta v|> v_{\rm esc}$. For absorbers with $|\Delta v|> v_{\rm esc}$, the median of $|\Delta v|$ is 650, 610, and 600 \kms\ but with median absolute deviations of 40, 50, and 100 \kms\ for low, intermediate, and highly ionized absorbers, again reflecting that the highest velocity absorbers are detected only in highly ionized species. Also notable is the comparative lack of $N_{\rm HI}>10^{14.5}$~\cm2 absorbers that are unbound (bottom panel). 

Clearly, the unbound CGM gas is multi-phase, similar to other gas within the CGM. The unbound metal absorbers typically appear to be associated with lower $N_{\rm HI}$, suggesting they are either highly ionized, metal rich, or both. 

The largest number of unbound absorbers is detected in \ion{C}{4} absorption. Among the 5 galaxies with detected unbound \ion{C}{4}, $\gtrsim 20\%$ of the \ion{C}{4} column density is unbound. Considering all of the detected \ion{C}{4}, $\gtrsim 12\%$ of the halo \ion{C}{4} column density is unbound. 

 Given the large velocity offset of the unbound absorbers as well as their high level of ionization and/or enrichment, the most plausible source of these absorbers is galactic winds - likely the same source as the \ion{O}{6} absorption discussed in \citet{tur15} and Section \ref{cf_OVI}. These data demonstrate that some of this gas, if it belongs to a galactic wind, has traveled large fractions of $R_{\rm vir}$ and still has sufficient velocity to escape the halo. The calculation above suggests a substantial fraction of the highly ionized halo gas may be unbound.

\section{The Temperature and Turbulence of Gas in the $z\sim2$ CGM}

\label{temp}

\begin{figure}
\center
\includegraphics[width=0.45\textwidth]{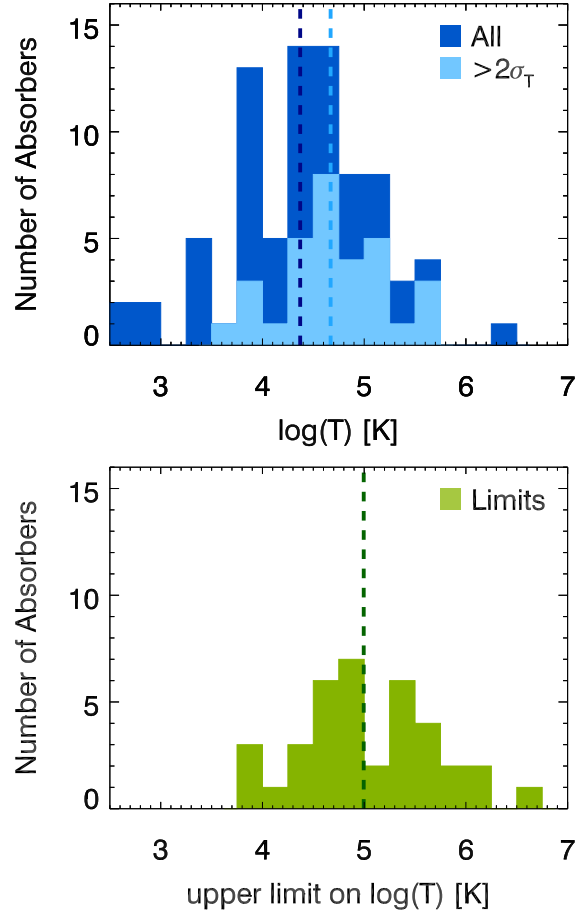}
\caption{The distribution of inferred gas temperatures and upper limits for the full sample of absorbers excluding those fit with pure turbulent broadening (see Section \ref{fits}). The upper panel shows the distribution of inferred gas temperature for the 80 absorbers with detected thermal broadening and inferred temperatures $T>10^{2.5}$~K. The darker shaded histogram show all of the measurements while the lighter-shaded histogram show absorbers with $>2 \sigma$ measurements of $T$. The median temperature for all of the multi-elements measurements is $\log (T / \rm K) =4.37$, shown in the dark blue vertical dashed line, while for the high-S/N sample, the median is $\log (T/ \rm K)=4.67$, shown in the light blue line. The lower panel shows the upper limits on the gas temperature for absorbers detected in only one element. The median value of these limits, shown by the vertical green line, is $\log(T/ \rm K )=5.00$. }
\label{T_lim}
\end{figure}

As described in Section \ref{fits}, in addition to providing kinematic information for gas within the CGM, the KBSS data also constrain the temperature and internal turbulence of gas within the halos of these galaxies. Within the last decade, much of the theoretical literature concerning the CGM has focused on the temperature and structure of accreting gas. The theoretical consensus is that most galaxies at early times do not form stable shocks at or near the virial radius of their dark matter halo because most have yet to build a massive virialized hot halo. And even after stable shocks have formed, for some dense filamentary gas configurations, much of the gas may still accrete without drastic heating \citep{bir03,ker05,ocv08,bro09,ben11,van11}. In these models, the filamentary structure of the cosmic web can penetrate the halos of even relatively massive (M$_{\rm{halo}}\approx 10^{12} \rm{M}_{\odot}$) galaxies, allowing efficient accretion of diffuse intergalactic material into the ISM of these early galaxies. The KBSS galaxies, with M$_{\rm{halo}}= 10^{12} \rm{M}_{\odot}$, provide a compelling test of this theoretical consensus. 

In \citet{gcr12}, we studied the Doppler width ($b_{ \rm d}$) distribution of \ion{H}{1} absorbers in the CGM of KBSS galaxies, finding that $b_{ \rm d,~HI}$ compared at fixed $N_{\rm{HI}}$ 
were higher close to galaxies than at random places in the IGM. These results suggested that gas within the CGM was either more turbulent or hotter. However, without the constraint provided by metal ions, we could not determine whether their broadening was due to higher gas temperatures or larger gas turbulence.  

The analysis presented herein includes a total of 130 detected metal-bearing absorbers, all within 100 pkpc impact parameter (roughly R$_{\rm{vir}}$) of a galaxy in the KBSS. With the additional analysis of metal absorption, we can address the source of line broadening within the gas. As discussed in Section \ref{fits}, differentiating between thermal and turbulent broadening of absorption systems requires the detection of at least two ions from elements with significantly different atomic masses (such as Si and C). Of the 130 absorbers, 93 (72\%) are detected in multiple metal ions covering at least two elements, making it possible to constrain their thermal and turbulent broadening. For all other absorbers, it is possible to place only upper limits on their gas temperature by assuming pure thermal broadening. Figure \ref{T_lim} shows the distribution of inferred gas temperatures for the full sample with multi-element measurements in the top panel and single-element absorbers with upper limits of the gas temperature shown in the bottom panel.\footnote{We note that all of the metal absorbers presented have associated hydrogen absorption; however, there are typically many metallic components whose associated hydrogen is completely blended. Thus, for the complex absorbers which seem to be characteristic of the CGM of $L^*$ galaxies at $z\sim2$, it is not possible to use the relative widths of \ion{H}{1} and some metallic transition to determine the gas temperature and turbulence.} For completeness, we also show the measurements with $T > 2\sigma_T$ in light blue in the upper panel. The three sets have median inferred temperatures of $\log(T / \rm K)$ of 4.37, 4.67, and 5.00 for all the detections, the high S/N sample, and the limits respectively. The median temperature for the entire sample (including upper limits) is $\log(T/ \rm K) = 4.5$. 

This temperature distribution is notable for several reasons. First, the gas spans $>2$ order of magnitude in temperature, even when only considering the highest significance measurements. Second, 40\% of the absorbers with measured gas temperatures (as well as 40\% of absorbers with upper limits) have inferred gas temperature between $4.5<\log(T/ \rm K)<5.5$ with an additional 6\% (10\% including limits) with $\log(T/ \rm K)>5.5$.

\begin{figure}
\center
\includegraphics[width=0.45\textwidth]{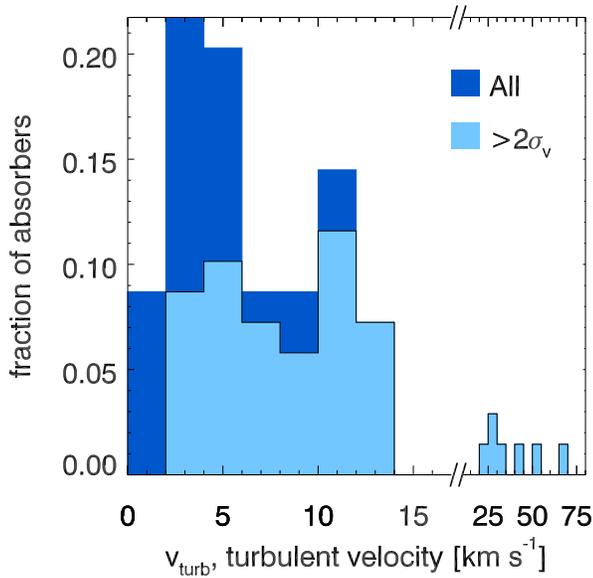}
\caption{Same as the top panel of Figure \ref{T_lim} but showing the turbulent velocity distribution of absorbers with detected turbulent broadening. The median turbulent velocity is $v_{\rm turb}=5.8$ \kms\ with an interquartile range of $3.7<v_{\rm turb}<10.7$ \kms. 90\% have $v_{\rm turb} < 15$ \kms.}
\label{turb_hist}
\end{figure}

\begin{figure}
\center
\includegraphics[width=0.45\textwidth]{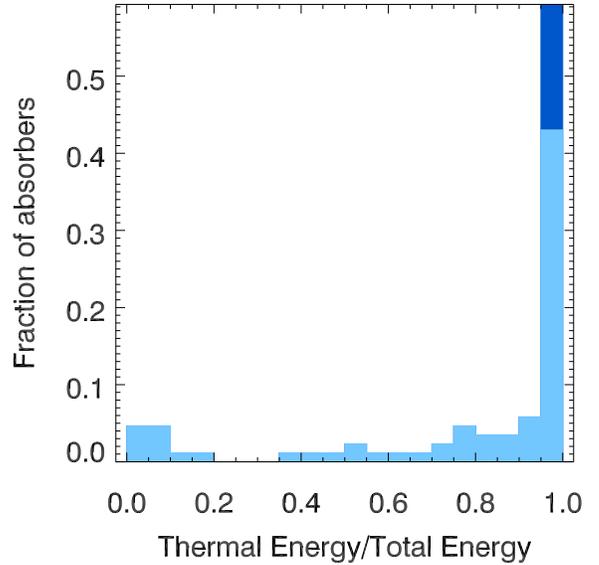}
\caption{ The fraction of the total internal kinetic energy attributed to thermal energy compared to the total kinetic energy from both non-thermal and thermal energy. The light blue histogram shows that a large fraction of the absorbers have much more thermal energy than kinetic energy from turbulence. The additional absorbers shown in dark blue are those fit with a pure thermal model. Notably, 58\% of absorbers fit with a thermal+turbulent model have $>90\%$ of their energy coming from thermal contributions while only 18\% have more turbulent than thermal energy.}
\label{energy_hist}
\end{figure}

In addition to constraining the gas temperature, the sample with multi-element detections can also be used to measure the internal velocity dispersion of individual gaseous structures, henceforth referred to as turbulence. Note that we do not measure the turbulence of the ensemble of `clouds' within the virial radius of a galaxy halo, but rather the dispersion of gas within a separable gaseous structure fit with a single Voigt profile.

Figure \ref{turb_hist} shows the measured turbulent velocities. Most of the studied gas in the inner CGM has relatively low measured turbulence. The median turbulent velocity is $v_{\rm turb}=5.8$ \kms\ with an interquartile range of $3.7<v_{\rm turb}<10.7$ \kms; 90\% have $v_{\rm turb} < 15$ \kms. 

\subsection{The Internal Energy of Halo Gas}

To compare the measured thermal and turbulent properties of the CGM, we compare the thermal and turbulent components of the internal energy of the gas. The energy per particle from the kinetic temperature of the gas is:
\begin{equation}
E_T=\frac{3}{2} kT
\end{equation}
This can be compared to the energy per particle due to the internal turbulence within the gaseous structures:
\begin{equation}
E_{\rm turb}=\frac{1}{2} \mu m_p v^2
\end{equation}
where $\mu=0.6$ is the assumed mean molecular weight, $m_{\rm p}$ is the proton mass, and $k$ is the Boltzmann constant.

Figure \ref{energy_hist} compares the thermal energy to the total internal energy in the gas. Notably, 58\% of absorbers fit with a thermal+turbulent model have $>90\%$ of their energy coming from thermal contributions while only 18\% have more turbulent than thermal energy. Clearly most of the internal energy in the gas comes from its temperature rather than bulk motions. 

\begin{figure}
\center
\includegraphics[width=0.5\textwidth]{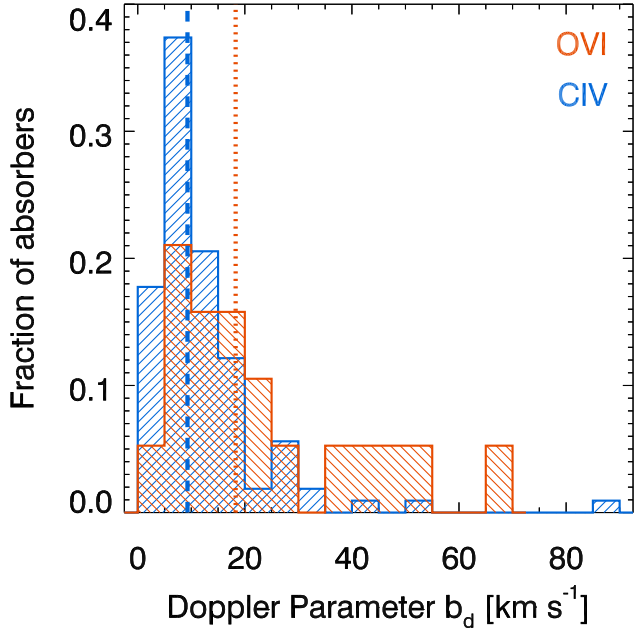}
\caption{The measured Doppler width of \ion{O}{6}-bearing absorption components (red) compared to \ion{C}{4}-bearing absorption components (blue). The median Doppler parameters of the samples are shown in the blue dashed (\ion{O}{6}) and red dotted (\ion{O}{6}) lines. Clearly the typical \ion{O}{6} absorber has significantly more thermal or non-thermal broadening than the typical \ion{C}{4} absorber. However, there exist a population of broad \ion{C}{4} absorption systems.   }
\label{b_CIV_OVI}
\end{figure}

\subsection{The Doppler widths of \ion{O}{6} absorbers }

\label{b_OVI}

The properties of \ion{O}{6}-bearing absorbers are of considerable interest given the expectation that \ion{O}{6} could trace a hotter gas phase, more similar to properties expected for gas heated by outflows or accretion shocks, cooling from $\log(T/ \rm K)>5.5$ \citep{hec02,for14,bor17,voi18}. 

The typical \ion{O}{6} line width in our sample is quite broad, and much broader than that of low and intermediate ions detected nearby in velocity (see Figure \ref{BX182}). The mean $b_{\rm d,~\rm{OVI}}=23.9$ \kms\ and a median $b_{\rm d,~\rm{OVI}}=18.3$ \kms\ compared with a mean \ion{C}{4} Doppler width of $b_{\rm d,~\rm{CIV}}=12.4$ \kms\ and a median $b_{\rm d,~\rm{CIV}}=9.3$ \kms. These values, if one assumes pure thermal broadening, correspond to \ion{O}{6} bearing gas with a mean (median) temperature of $\log(T/ \rm K)=5.7~ (5.5)$ and \ion{C}{4} bearing gas with $\log(T /\rm K)=5.0 ~(4.8)$. Notably, for \ion{O}{6} and \ion{C}{4} a temperature  $\log (T / \rm K)=5.5$ and $\log(T/ \rm K)=5.0$ corresponds to temperature at which the O$^{5+}$/O and C$^{3+}$/C ratios peak for gas in collisional ionization equilibrium.

\begin{figure}
\center
\includegraphics[width=0.5\textwidth]{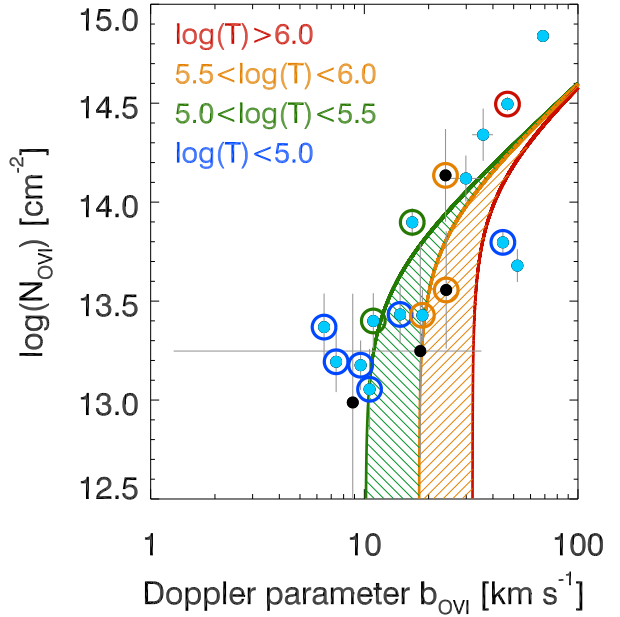}
\caption{The measured Doppler width of \ion{O}{6}-bearing absorption components compared to their column densities. Absorbers plotted in light blue are $>2\sigma$ significant measurements of $N$. A Spearman-Rho test confirms that $N_{\rm OVI}$ is positively correlated with  Doppler width at 4.3$\sigma$ significance. Absorbers with measured temperatures are surrounded by circles whose color indicates the inferred gas temperatures. The gas temperature is also positively correlated with column density, although the significance of this correlation is lower. The curves and shading show the cooling models from \citet{hec02} who predict a relationship between $N_{\rm OVI}$ and $b$ for shock-heated radiatively cooling gas with $5\lesssim \log(T/\rm K)\lesssim 6$. Note that here we assume the \citet{hec02} curves represent the width of \textit{individual} absorbers, not the total velocity width of all \ion{O}{6}. }
\label{b_N_OVI}
\end{figure}

Figure \ref{b_CIV_OVI} compares the full distribution of $b_{\rm d}$ for \ion{O}{6} and \ion{C}{4} absorbers. It is evident from both the bulk statistics and the full distribution that the \ion{O}{6} gas has significantly larger Doppler parameters; however, it is of interest that there exist a sample of broad \ion{C}{4} absorbers as well. The large line widths and, in some cases, corresponding high temperatures suggest that many of these broad absorbers, most commonly detected in \ion{O}{6} and \ion{C}{4}, are collisionally ionized and consistent with the properties of a metal-enriched hot halo or hot gas outflow. Future ionization analysis will more fully address the origin of this gas.

Previous absorber studies have found similar trends in other environments. \citet{fox07b,fox07a} studied the properties of \ion{C}{4} and \ion{O}{6} absorbers associated with high column-density \ion{H}{1} absorbers (DLAs and sub-DLAs). They also found both narrow and broad \ion{C}{4} components and broad \ion{O}{6} absorbers they suggest were collisionally ionized. \citet{muz12} considered the full population of \ion{O}{6} absorbers in QSO spectra and found that the Doppler parameters of \ion{O}{6} absorbers were typically larger than those of \ion{C}{4}. \citet{sim02} measured the widths of \ion{O}{6} absorbers in high-$z$ QSO spectra, finding a median $b_{\rm d,~\rm OVI}=16$ \kms, similar to the value we find here.

Figure \ref{b_N_OVI} shows the measured column densities, line widths, and temperatures of \ion{O}{6} absorbers in our sample. Notably both the widths of \ion{O}{6} absorbers and their inferred temperatures are correlated with their column densities.  A Spearman-Rho test confirms that $b_{\rm d}$ and $N$ are significantly correlated ($4.3\sigma$). The correlation between $N$ and $b_{\rm d}$ is stronger than the correlation between $N$ and $T$, even if the sample is restricted to the fraction of the sample with temperature measurements. 

The correlation between $b_{ \rm d}$ and $N_{\rm OVI}$ has been noted previously at lower redshift (see e.g. \citealt{hec02,wer16,bor17}.) \citet{hec02} and more recently \citet{bor17} have proposed that such a correlation would be expected in the case of $T\sim10^{5-6}$ K gas radiatively cooling behind a fast shock. The data here generally agree with the predictions from \citet{hec02} when one considers the  Doppler widths of \textit{individual absorbers} rather than the velocity width of all the \ion{O}{6} absorption. If this model accurately reflects the source of \ion{O}{6} absorbers, it is notable that the model assumes metallicities $Z/Z_\odot > 0.1$ suggesting the gas detected in \ion{O}{6} is typically quite enriched. 

Given the generally high inferred temperatures of the \ion{O}{6} absorbers, we do not compare directly to the models proposed by \citet{ste18} where the gas is photoionized; however, we note that the absorbers with lower inferred temperatures may be better described by such a model (see e.g. \citealt{zah18}). Alternatively, absorbers with low inferred temperatures could be over-ionized gas that has rapidly cooled. For initially hot gas radiatively cooling to $T\lesssim10^6$ K, non-equilibrium ionization effects are likely due to the short cooling times at these temperatures. Rapid radiative cooling leads to the recombination of the gas lagging behind the drop in temperature resulting in over-ionization of the gas compared to collisional ionization equilibrium \citep{gna07, opp13, gna17}. 

Further analysis of this \ion{O}{6}-bearing gas in warranted. Any scenario attempting to explain the origin of these absorbers must explain not just their measured $N_{ \rm OVI}$, but also their $b_{\rm d}$ and measured temperatures. Future work will more fully consider the ionization conditions within this sample which may further elucidate the origin of this gas.

\section{Discussion}

\subsection{The Implications of the Measured Gas Temperatures}

\label{temp_discuss}

As discussed in Section \ref{temp}, the thermal properties of the gas within the CGM offer direct insight into the properties of accreting and outflowing gas, constraining the physical processes affecting gas accretion and those driving galactic winds. Because of this, the temperature distribution shown in Figure \ref{T_lim} is of considerable interest. 

The inferred temperature range of the halo gas is not unexpected given the ionization states we use to trace the CGM, which for collisionally ionized gas peak in the temperature range of $4<\log{(T/ \rm K)}<5.5$. However, the range of temperatures found is notable for a number of reasons. First, the measurements require that much of the gas within the halos of $L^*$ galaxies at $z\sim2$ has temperatures at which radiative cooling is most efficient \citep{dal72}. For such gas, the cooling time is short.\footnote{The precise cooling time of gas is determined by its density and metallicity; however, for all gas with densities higher than the mean density at $z\sim2$, the cooling time for $T=10^5$K is less than $10^8$ years. For the somewhat higher densities expected in the CGM, and for higher metallicities, the cooling time decreases substantially.}  Given this, gas at these temperatures is expected to be short lived unless it is subject to nearly continuous heating. Alternatively, the existence of intermediate temperature gas could be common if such gas forms via cooling out of a more massive hot phase. \citep{mal04,mcc12,voi17,sca17,sch18,gro18,sur18}

The majority of the gas appears to be hotter than the typical temperature of hydrogen clouds within the $z\sim2$ IGM. Inferences based on \ion{H}{1} absorbers indicate that the IGM has a typical temperature at mean density of $\sim 1-2 \times 10^4 $~K \citep{hui97,sch99, gcr12b}. More than half of the absorbers in our sample have line widths broader than expected for that temperature. 

The IGM temperature is set by a balance between photoionization heating and adiabatic cooling due to the expansion of the universe. Within a galaxy's dark matter halo, the gas would not be freely expanding, so significant adiabatic cooling would be unlikely. Therefore, the equilibrium temperature in the absence of accretion shocks or heating from galactic winds would be set by the balance of photoionization heating and radiative cooling. Around the mean cosmic density at $z\sim2$, this equilibrium temperature is insensitive to metallicity and reaches as high as $2\times 10^5$ K \citep{smi17}. However, the expected gas density within galaxy halos is significantly higher than the mean density of the universe. At higher densities, the equilibrium temperature drops such that at 100 times the mean density (still lower than the typical dark matter density within the halo), $T_{\rm equ} \approx 5 \times 10^4$ K with only a modest dependence on metallicity.

While much of the gas is hotter than random places in the IGM and hotter than expected for photoionization heating, the temperature of almost all of the gas we detect is lower than the virial temperature of the halo in which it resides. The virial temperature of a dark matter halo of  virial radius $R_{\rm vir}$ and halo mass $M_{\rm halo}$ is defined to be: 
\begin{equation}
T_{\rm vir} = \frac{\mu m_{\rm p} GM_{\rm halo} }{2R_{\rm vir}k}
\end{equation}
where $\mu=0.6$ is the assumed mean molecular weight, $m_{\rm p}$ is the proton mass, and $k$ is the Boltzmann constant. Assuming the typical halo mass and virial radius for the sample, $M_{\rm halo}=10^{11.9} M_{\odot}$ and $R_{\rm vir}=90$ kpc \citep{tra12}, the typical virial temperature for the galaxies in this sample is $T_{\rm vir} \approx 10^6$ K. 

At $T\sim10^6$ K, we expect collisional ionization to be dominant and  \ion{C}{4} and \ion{Si}{4} to be poor tracers of the gas since most C and Si would be more highly ionized. Even \ion{O}{6}, which has the highest ionization potential of any of the ions we can currently measure, peaks in collisional ionization equilibrium at $T=10^{5.5}$ K. Thus, these measurements do not preclude the existence of gas with $T \gtrsim 10^{6}$ K within the halos of these galaxies. However, the data do require a substantial reservoir of gas at intermediate temperatures. 

The data presented are not at odds with the picture of cold flows of filamentary accretion. Roughly half of the absorbers are cool with $T=10^4-10^{4.5}$K. However, given that much of the gas is hotter than is typical within the IGM, it is clear that additional heating is present. The intermediate temperature of much of the gas within the CGM is concrete evidence that the gas is affected by its proximity to the galaxy, most likely through shock heating during accretion or via galactic winds, with much of the intermediate temperature gas likely resulting from cooling. 

As the thermal state of CGM gas is directly tied to the physics of outflows and/or accretion shocks, it is likely that these measurements are a powerful diagnostic which should be used to constrain physical models implemented in hydrodynamic simulations. Ideally, such simulations would report the temperature distribution of gas weighted by its ionic column densities, or other analogous quantities that can be compared directly with this data set. In the absence of such predictions, we compare here to temperature distributions made for the mass of gas and metals within the halo of simulated galaxies. Given that we do not measure directly the mass of the gas or metals in these observations, these comparisons should be viewed as exploratory rather than conclusive. 

\citet{mur17} and \citet{haf18} analyzed the temperature distribution of gas within $R_{\rm vir}$ of galaxies at $z=2-2.5$ in the FIRE simulations. \citet{mur17} report the fraction of metal mass as a function of temperature in three simulations with stellar masses $M_{*}=10^{9.5} M_\odot$ at $z>2$. They find less than 20-30\% of the metal mass has $T>10^{4.7}$ K, with roughly 60\% of the metals having  $10^4 < T<10^{4.7}$ K. Considering galaxies with halo masses $M_{\rm halo}=10^{11.5} M_\odot$, \citet{haf18} found that 40-50\% of the total CGM gas mass (including hydrogen) is cool [$4 < \log{(T/ \rm K)} < 4.7$] and another 40-50\% is hot [$\log{(T/ \rm K)} > 5.3$] with less than 10-20\% of the gas having warm gas temperatures [$4.7 < \log{(T/ \rm K)} < 5.3$]. 

While we can not constrain the \textit{mass fraction} of gas as a function of temperature, over half (61\% by number) of the absorbers in our sample have $\log{(T/ \rm K)} < 4.7$; 20\% of absorbers in our sample with measured temperatures would be classified as warm [$4.7 < \log{(T/ \rm K)} < 5.3$],\footnote{If we include absorbers with upper limits on their temperature, this fraction is unchanged.} and 17\% of the absorbers have line widths or measured temperatures consistent with hot gas [$\log{(T/ \rm K)} > 5.3$]. 

The measured KBSS values \textit{weighted by number of absorbers} are similar to those presented by \citet{mur17} averaging by metal mass, but there appears to be a somewhat larger fraction of gas with $\log{(T/ \rm K)} > 4.7$ in the KBSS galaxy halos than was found by \citet{mur17}. Alternatively, presuming that we are missing some fraction of the hot absorbers because they do not produce detectable absorption in the observed ions, the observed temperature distribution of absorbers \textit{weighted by number} appears roughly consistent with the \textit{mass-weighted} fractions reported by \citet{haf18}. Given that it is unlikely that the mass-weighted and number-weighted temperature distributions are the same, additional analysis of the simulations may allow for better comparisons with the data in the future.

\citet{fie17} considered idealized high-resolution simulations of galactic halos to study the thermal properties of gas around galaxies and how they are affected by inflows and outflows. Depending on the details of the feedback efficiency used, the simulations showed significant differences in the quantity of gas at intermediate temperatures [$4.5<\log({T/ \rm K})<5.5$]. Importantly, the halos that showed this effect were $\sim0.5$ dex lower mass than the halos considered here; however, the higher mass halos showed less variation due to the predominance of thermal support from a hot halo. One caveat in the comparison is that the simulations were initialized with a hot halo of gas at $T=T_{\rm {vir}}$, which is unstable at $\log{(M_{\rm{halo}}/M_\odot)}<11.5$, but stable for galaxies in the mass range of the KBSS. However, if the KBSS galaxies have not yet built a stable and massive hot halo, perhaps the intermediate temperature gas in KBSS halos is also particularly sensitive to feedback. If so, the temperature distribution of KBSS halo gas may be a critical probe of properties of galactic outflows within galaxies in the distant universe. This possibility should be investigated further. 

Recently, several studies have focused on the existence of cool ($T\sim10^4$ K) and warm ($T\sim10^5$ K) gas in galactic winds. The measurements here provide a critical test of these outflow models. In most detailed models, galaxies drive out hot ($T>10^6$ K) gas, rapidly shredding any embedded cold clouds via Kelvin-Helmholz instabilities.  However, in some models, comoving cold gas is reformed via radiative cooling instabilities \citep{tho16,sca17,sch18, gro18}. Another class of models drive  galactic winds via cosmic ray pressure. Recent theoretical work has shown that winds driven by cosmic rays are cooler \citep{boo13,gir18} and that the overall mass with $T<10^5$ K in the low-redshift CGM is substantially increased \citep{sal16}. 
\citet{sal16} and \citet{but18} showed that the thermal structure of the CGM is highly sensitive to both the mode and detailed treatment of cosmic ray transport (advective streaming vs isotropic or anisotropic diffusion). Clearly, the data presented herein can test or constrain many aspects of these models; however, such a comparison requires detailed predictions from simulations that are matched to the properties we can observationally constrain (column densities of observed ions, ion-weighted temperatures, projected velocities). More detailed comparisons should be made between these and other outflow models and these observations. 

\begin{figure}
\center
\includegraphics[width=0.40\textwidth]{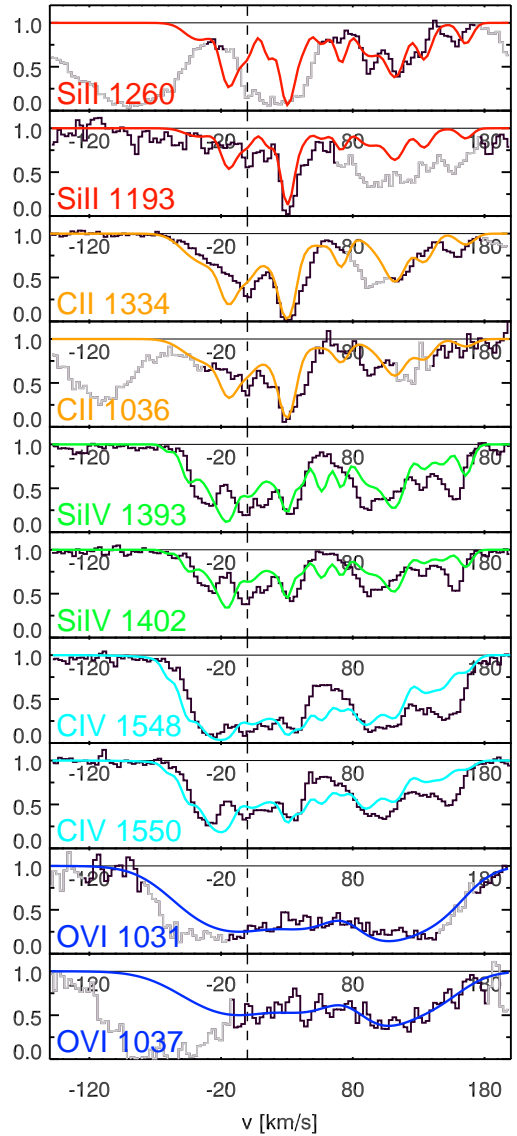}
\caption{Constraints on the size scale of absorbers within the halo of Q0142-BX182 from a lensed background QSO. The colored curve shows the Voigt profile fit to the brighter image (UM673A) of Q0142. The black spectrum is the continuum-normalized spectrum of the second image (UM673B) of the QSO, which probes the halo gas 400 pc from the brighter image. Notably, the overall velocity spread of the gas is nearly identical across the two sightlines suggesting the metal enriched structure giving rise to these absorbers spans distances $\gg 400$ pc. Similarly, the \ion{O}{6} spectrum appears unchanged across this distance range; however, the detailed structure within all of the lower-ionization gas (including \ion{C}{4}) has significant differences between the two sightlines, suggesting individual clouds or density inhomogeneities within the halo typically have sizes $<400$ pc.  As in Figure \ref{BX182}, the grey sections of the spectrum show regions of contamination within the spectrum.}
\label{BX182_imageb}
\end{figure}

\subsection{Constraints on the sizes of absorbers within the halo}

\label{size}


Another interesting consideration is the low measured turbulent broadening of the gas. The distribution in Figure \ref{turb_hist} provides useful constraints for theoretical models and simulations attempting to explain or reproduce observations of the CGM. In particular, many recent theoretical studies have focused on the small-scale structure within the CGM, suggesting that the CGM may be made of up very small clouds, more analogous to a terrestrial fog or mist \citep{mcc18,voi17,hum18, lia18}. Depending on the true size scale of such ``cloudlets'' and whether they are smaller or larger than the projected size of the background QSO at the location of the cloud (sizes as small, or smaller than, $10^{-4} - 10^{-2}$ pc \citealt{mor10,dai10,jim14,mud18}), each component of our Voigt profile fit corresponds either to a single cloudlet or a coherent aggregate of such structures. In the former scenario, these observations constrain the turbulence of gas within single cloudlets. In the latter, they represent the velocity dispersion of the cloudlets within a single kinematically coherent structure, and the number of detected components gives an indication of the volume filling of such structures. 

Our measurements suggest that internal turbulence within coherent gas structures in the halo is sub-dominant energetically compared to the thermal energy for over 80\% of the metal-bearing absorbers. This limits the number of cloudlets that can be contained within a single absorber fit with one Voigt profile \citep{mcc18}. 

However, the data also do not show any evidence of partial covering of the QSO beam. Multiple transitions in various galaxy halos are saturated, with absorption lines reaching zero transmission suggesting no light passes from the background QSO unattenuated. Similarly, there is no evidence for doublet ratios inconsistent with the values set by their atomic parameters, as would be expected if only part of the QSO beam were intercepting gas along the line of sight. Given these observations, if the cloudlets are smaller in size than the QSO beam, a sufficiently large number must always be created to cause 100\% areal covering fractions within the coherent gas structure, even if the volume filling factor is substantially lower.

Within the KBSS sample, we have one unambiguous constraint on the size scale of metal absorbers. One of the QSOs in our sample, Q0142 (UM673) is lensed by a foreground $z=0.49$ galaxy \citep{sur87} producing two images separated by 2.2 arcseconds, one roughly two magnitudes fainter than the other \citep{sme92}. We obtained Keck/HIRES spectra of both images, as discussed in \citet{coo10} as well as data drawn from the Keck archive and discussed in \citet{rau01}.
These two spectra constrain the coherence of halo absorbers at 75 kpc from one of the $z>2$ KBSS galaxies, BX182. Following the calculations in \citet{coo10}, we derive a separation of the two sightlines of $\sim 400$ pc at the redshift of Q0142-BX182. We note that these spectra were analyzed by \citet{rau01} to study the coherence scale of IGM metal absorbers (see their Figure 1, right hand panel). Here we only consider those absorbers within $\pm1000$ \kms\ and 100 kpc of Q0142-BX182.

Figure \ref{BX182_imageb} shows a comparison of the Voigt profile fit to the spectrum of UM673A (colored curves) compared to the spectrum of UM673B (black spectrum) which probes the gas 400 pc from the A sightline. Overall, the velocity range covered by metal enriched gas is constant across the two sightlines suggesting the metal enriched structure responsible for the metal absorption within the halo of Q0142-BX182 is larger than 400 pc. The \ion{O}{6} section of the spectrum appears completely consistent between the two sightlines, suggesting there are no significant differences in velocity or density within the \ion{O}{6}-bearing phase on such small scales.

Conversely, the detailed subcomponent structure in the lower-ionization gas (including \ion{C}{4}) shows significant differences between the two lines of sight: the strengths, widths, and centroids of the various components are typically different between the two sightlines. There are some absorbers, such as the strong low-ionization absorber at $v\sim30$ \kms, that appear unchanged between the two sightlines; however most have noticeable differences. This suggests that the majority of halo clouds (or coherent cloudlet structures) are smaller than (or have density variations on scales comparable to) the sightline separation, 400 pc, while the overall structure containing metal enriched gas within the halo is much larger, consistent with the conclusions of other absorber studies with lensed QSOs \citep{rau01,ade05,che14,zah16}. Further, these patterns imply that the \ion{O}{6}-bearing gas likely is more volume-filling than the lower-ionization gas. 

\subsection{Comparison of the derived temperature and gas turbulence to low-redshift CGM measurements}

\citet{zah18} compared the Doppler parameters of \ion{Mg}{2} and \ion{H}{1} absorbers within the inner CGM of massive elliptical galaxies at $z\sim0.4$. These authors report that the low ionization CGM has a mean temperature of $2 \times 10^4$ K, similar to typical IGM temperatures and cooler than we have found here.  The measured turbulence within these \ion{Mg}{2} absorbers was similar to that found in our sample with $v_{\rm turb}=7\pm 5$ \kms. 

Similar to \citet{gcr12}, \citet{tum13} used the measured Doppler widths of \ion{H}{1} absorbers in the CGM of galaxies in the COS-Halos survey to measure an upper limit on the temperature of \ion{H}{1} bearing gas of $\log(T/\rm K)\lesssim5.3$ K assuming pure thermal broadening.  Compared to IGM samples of \ion{H}{1} absorbers selected without knowledge of the positions of galaxies \citep{tri08,tho08,til12}, \citet{tum13} found that the CGM $b$-parameter distribution was consistent with that of the full IGM, i.e. they do not find that HI absorbers in the low-$z$ CGM are characteristically different from the full distribution of IGM absorbers, even for gas surrounding the massive early-type population \citep{tho12}.  

Both \citet{tho12} and \citet{tum13} note that the inferred gas temperature is less than the virial temperatures of the expected dark matter halos of the COS-Halos galaxies; however, they note that this result is unsurprising given that the tracer is neutral (\ion{H}{1}) and that broad absorbers consistent with million degree gas are very challenging to detect, especially when blended with the narrower \ion{H}{1} absorption commonly found surrounding COS-Halos galaxies. 

\subsection{Comparison of the derived temperature and gas turbulence to other measurements in the IGM}

\label{IGM_discuss}

The measurements presented herein are not the first constraints on the temperature and turbulence of gas in the distant universe. Several studies have focused on the thermal and turbulent properties of absorbers in high-$z$ QSO sightlines, henceforth referred to as ``IGM'' absorbers. Importantly, the analyzed absorbers were detected in ionized metal species (a requirement since the ratio of the widths of those lines is used to measure the temperature and non-thermal motions in the gas). Given that metals originate within stars and galaxies, these absorbers do trace a biased sample compared to general \ion{H}{1} absorbers. Without knowledge of the galaxy distribution surrounding these ``IGM'' absorbers, it is not possible to say what fraction of them are actually within the CGM of a galaxy like the ones in our sample, or any galaxy. But it is important to recognize the likelihood that they do not, as a sample, represent completely random locations within the IGM. 

The first measurements of the thermal and non-thermal broadening measurements for a large sample of $z\sim3$ IGM absorbers were reported in \citet{rau96}. These authors considered a sample of 79 \ion{C}{4} absorbers with detected \ion{Si}{4} and found a mean $b_{\rm d,~\rm{CIV}}=12.6$ \kms and a median $b_{\rm d,~\rm{CIV}}=9.6$ \kms.  In our CGM sample, we find a mean $b_{\rm d,~\rm{CIV}}=12.4$ \kms and a median $b_{\rm d,~\rm{CIV}}=9.3$ \kms in complete agreement with the values reported by \citet{rau96}. Decomposing the Doppler widths into thermal and non-thermal components, \citet{rau96} found a typical $v_{\rm turb}=6.3$ \kms, slightly smaller but consistent with our derived measure of turbulence. Similarly, they derive a typical temperature of $4\times 10^4$K, consistent with the median values shown in Figure \ref{T_lim}. 

More recently, \citet{kim16} used a sample of 54 kinematically simple \ion{C}{4} and \ion{C}{3} $ 2.1 < z < 3.4 $ IGM absorbers for which the associated \ion{H}{1} absorption could be isolated. By comparing the $b_{\rm d}$ parameters of the C and H absorbers, these authors derive gas temperatures with a mean of $\langle \log{(T / \rm K)} \rangle = 4.27 \pm 1.0$ and median $\log{(T / \rm K)} = 4.47$, again consistent for the values found within the CGM. 

Both of these comparisons imply significant similarities between the typical metal-bearing absorber selected at random from a QSO line of sight (``IGM'' absorber) and the typical metal-bearing absorber known to be within the CGM of a galaxy $z>2$. While the temperature of gas within the CGM typically appears to be higher than random \ion{H}{1} absorbers within the IGM, it is similar to that of metal-bearing IGM gas. Given that all metals within the IGM must have once passed through or still be contained within the CGM of some galaxy, this conclusion is not surprising. However, given the very rapid cooling time of gas at the intermediate temperatures that compose half of the detected gas, this may suggest that a non-negligible fraction of IGM absorbers at $z\sim2-3$ have been ejected from or in some other way heated by a galaxy in the recent past. 

\begin{deluxetable}{lccccc}
\tablecaption{Metal Mass within the Halo}  
\tablewidth{0pt}
\tablehead{
\colhead{Ion} & \colhead{min($\log(N_x)$)\tablenotemark{a}} & \colhead{$D_{\rm max}$\tablenotemark{b}} & \colhead{$f_c$} & \colhead{$\log(\langle N \rangle)$\tablenotemark{c}} & \colhead{$\log($M$_x)$}\\
\colhead{} & \colhead{[\cm2]} & \colhead{[pkpc]} & \colhead{} & \colhead{[\cm2]} & \colhead{[M$_\odot$]}}
\startdata
\ion{Si}{2}	  &   12.6	&      75     &	     0.6  &   13.85   &	  5.2   \\
\ion{C}{2}	  &   12.4	&      75     &	     0.6  &   14.46   &	  5.5	\\
\ion{Si}{3}	  &   \nodata	&      90     &	     1.0  &   13.23   &	  5.0	\\
\ion{C}{3}	  &   \nodata	&      90     &	     1.0  &   14.20   &	  5.6	\\
\ion{Si}{4}	  &   \nodata	&      90     &	     1.0  &   13.62   &	  5.4	\\
\ion{C}{4}	  &   \nodata	&      90     &	     1.0  &   14.31   &	  5.7	\\
\hline
\ion{O}{6}    &   \nodata	&      75     &	     1.0  &   14.80   &	  6.2   \\				\hline			  
\hline
Total C  &   \nodata	&   \nodata   &	\nodata	  & \nodata   &	  6.1	\\
Total Si &   \nodata	&   \nodata   &	\nodata	  & \nodata   &	  5.7	\\
Total O \tablenotemark{d}	 &   \nodata	&   \nodata   &	\nodata	  & \nodata   &	 6.9 \enddata
\tablenotetext{a}{For ions with non-detections within $D_{\rm max}$, the minimum column density detectable in all sightlines.}
\tablenotetext{b}{The impact parameter of the farthest galaxy in the sample that has a detection of the ion.}
\tablenotetext{c}{The mean column density among the detections.}
\tablenotetext{d}{The total Oxygen mass calculated assuming 20\% of oxygen is in the state O$^{5+}$.}
\label{mass_halo}
\end{deluxetable}

\subsection{The Metal Mass in the High-Redshift CGM}

\label{mass_discuss}

We close with a rough estimate of the metal mass detected within the halo of KBSS galaxies.
We calculate the implied mass in metals using the total column densities, $\Sigma N$, and covering fractions $f_c$ measured in Section \ref{cf_section}. We begin by calculating the measured mass in each ion:
\begin{equation}
M_{\rm ion} = \pi D_{\rm max}^2 m_{\rm ion} f_c \langle N_x \rangle
\end{equation}
where $D_{\rm max}$ is the maximum impact parameter at which the ion is measured in the sample, $m_{\rm ion}$ is the mass of the ion, $f_c$, is the covering fraction of the ion measured within $D_{\rm max}$, and $\langle N_x \rangle$ is the average column density measured across the detected sample. We note that for \ion{C}{3}, \ion{Si}{3}, \ion{Si}{4} and \ion{C}{4} many of the measured $\Sigma N$ are lower limits. We treat them as detections in the average, and so the values reported here are lower limits on the mass within the halo.

Table \ref{mass_halo} shows the measured values for these parameters and the implied mass in each ion within $R_{\rm vir}$ which corresponds to $10^{5-6}$ $M_\odot$ for each ion. For photoionized gas, we would expect to detect the majority of the phases of Si and C within this sample, so we sum the measured \ion{Si}{2}, \ion{Si}{3} and \ion{Si}{4} as well as the measured \ion{C}{2}, \ion{C}{3} and \ion{C}{4} to find the total metal mass within the photoionized phase: $M_{\rm Si, photo} \ge 10^{5.7}$ $M_\odot$ and $M_{\rm C, photo} \ge 10^{6.1}$ $M_\odot$. For oxygen, we only detect the highly ionized \ion{O}{6} stage. \ion{O}{6} traces $<20\%$ of the oxygen, regardless of the ionization mechanism \citep{tum11,opp13}, so the total oxygen within the phase traced by \ion{O}{6} is at least 5 times larger: $M_{\rm O, hot}\ge 10^{6.9}$ $M_\odot$. 

To place these values into context, we compare to the total mass in each ion with the mass of that element we expect to have been produced by a typical KBSS galaxy as well as the observed mass of metals contained within the ISM and locked in stars as shown in Table \ref{mass_in_gal}. 

We begin with an estimate of the total amount of metals produced by the galaxy. The IMF weighted yields of a stellar population carry significant uncertainties. These are driven by uncertainties in the IMF itself and by assumptions about the highest mass stars which contribute to nucleosynthesis and to a lesser extent, the effect of rotation and mass loss in massive stars, and uncertainties in the yields of individual supernovae \citep{vin16}.
These uncertainties propagate into nearly an order of magnitude uncertainty in the total yield of a given stellar population. Reasonable assumptions can lead to values that range from 0.006 $M_\odot$ of oxygen for every $M_\odot$ of stars formed \citep{nom06} to upwards of 0.03 $M_\odot$ and potentially even higher from top-heavy IMFs \citep{vin16}. Given the large uncertainty, we choose as a fiducial value 0.015 $M_\odot$ of oxygen for every solar mass formed, as this value has commonly been assumed in previous CGM studies \citep{tum11,pee14,joh17}, but we caution again that this value is highly uncertain.

\begin{deluxetable}{lccc}
\tablecaption{Metal Mass in Galaxies \label{mass_in_gal}
}  
\tablewidth{0pt}
\tablehead{
\colhead{Location} & \colhead{$\log(M_{\rm C})$} & \colhead{$\log(M_{\rm Si})$}  & \colhead{$\log(M_{\rm O})$} \\
\colhead{} & \colhead{[$M_\odot$]} & \colhead{[$M_\odot$]} & \colhead{[$M_\odot$]} }
\startdata
Photoionized Halo gas & >6.1 & >5.7 & \nodata  \\
\ion{O}{6}-Bearing Halo Gas & \nodata & \nodata & >6.9  \\
ISM &     7.1  &   6.7  &   7.7	\\
Locked in stars & 7.2 & 6.8 & 7.8 \\
Produced by the stars &  8.0  &   7.6  &   8.6 
\enddata
\end{deluxetable}

The mean stellar mass of the KBSS galaxies in this sample is $\langle M_* \rangle=10^{10.3} $ M$_\odot$ \citep{str17}. For a 100 Myr old stellar population, the total mass of stars formed is roughly 1.4 times the current stellar mass\footnote{After 100 Myr, $(1.4-1.0)/1.4\approx 30\%$ of the mass in the stellar population has been returned to the ISM via stellar winds or supernova ejecta.} for a \citet{chabrier} IMF \citep{bc03}. Thus, we expect a typical galaxy in our sample to have formed $10^{10.4} $ M$_\odot$ of stars and $10^{8.6} $ M$_\odot$ of oxygen. Using either the solar abundances from \citet{asplund} or the core-collapse supernova yields from \citet{woo07} or \citet{nom06}, the Si/O mass fraction is  $\sim0.1$, implying $10^{7.6} $ M$_\odot$ of Si. The nucleosynthetic origin of carbon is more complex, and C/O mass fraction range from $\sim0.1-0.5$. We select C/O$=0.25$ as it is in the middle of this range, and is the value measured within ionized HII regions within KBSS galaxies \citep{ccs16}. Using this value implies an assembled mass of $10^{8.0} $ M$_\odot$ of C.

This suggests that within the halo we have only detected  $\sim 1-2\%$ of the O, C, and Si formed by the stellar population, although with large uncertainties dominated by uncertainties in the IMF-weighted yields. 

We can also compare the halo metal mass to an estimate of the metal mass found within the ISM of these galaxies. \citet{tac10} measured the gas mass in typical star-forming galaxies at $\langle z \rangle=2.3$, finding that the cold ISM contained 44\% of the total baryonic mass within the galaxy [$M_{\rm gas}/(M_{\rm gas}+M_{\rm *})=0.44$], implying $M_{\rm ISM}=10^{10.2} $ M$_\odot$ for the KBSS galaxies in our sample. The abundance ratios within the molecular gas are not well constrained by the \citet{tac10} data, so instead we rely on abundances derived from nebular emission from HII regions. Since the massive stars are formed out of molecular gas within these galaxies, we assume for this estimate that the abundances are similar in the ionized, neutral, and molecular phases. \citet{str18} report that KBSS galaxies have a median gas-phase oxygen abundance $12+\log(\rm O/\rm H)=8.37$\footnote{Note that this is a fraction by number, not by mass.}, implying an oxygen mass within the ISM of $M_{\rm O, ISM} =10^{7.7}$ M$_\odot$. We assume the same abundance ratios as were discussed above to estimate the carbon and silicon ISM masses, finding $M_{\rm C, ISM} =10^{7.1}$ M$_\odot$ and $M_{\rm Si, ISM} =10^{6.7} $ M$_\odot$. Comparing these values with the total yields calculated above suggest that the ISM of KBSS galaxies has retained $\sim10\%$ of the metals formed.

Rounding out the census of observed metals are those locked within stars at the time of observation. To estimate the metal mass in stars, we assume all stars in the galaxy have the same abundance as the ISM.  We use the same abundance ratios as above and the average stellar mass to calculate $M_{\rm O, star} =10^{7.8} $ M$_\odot$, $M_{\rm C, star} =10^{7.2} $ M$_\odot$ and $M_{\rm Si, star} =10^{6.8} $ M$_\odot$.

Our lower limit on the total oxygen in the highly ionized \ion{O}{6} phase within 75 kpc is 15\% of the ISM oxygen. The lower limit on the carbon and silicon in the photoionized phase is 10\% of the ISM carbon and silicon mass. The kinematic structure of the various ions imply that \ion{O}{6} traces a distinct phase of gas compared to the likely photoionized Si and C, therefore, the masses that they trace are also distinct. If we assume the abundance pattern is similar in the photoionized and \ion{O}{6}-bearing phase, we can add their ISM fractions, suggesting we are tracing a total metal mass within the halo equivalent to $>25\%$ of the ISM metal mass.

While the detected halo metal mass is sizable, these calculations suggest we may have yet to account for the majority of the metals formed by the galaxy. Given the uncertainties in the total metal yields discussed above, it is plausible that a factor of a few or larger error exists in these calculations, and that the majority of the metals produced are actually accounted for. However, it is also possible that a large fraction of the metals are yet to be observed. If so, where are these metals? 

The data analyzed here are not particularly sensitive to hot gas with $T \gtrsim 10^6$ K, and so it is possible that a significant fraction of the metals are too hot to be studied via UV absorption. Further, $\sim90\%$ of our sample is probed at $D_{\rm tran} > 0.5 R_{\rm vir}$, with only one low-mass galaxy probed at smaller separations. Therefore, it is plausible that the column densities of metals are larger within the inner CGM, allowing for a substantial fraction of the metals to still lie within the halo. 

An additional possibility, however, is that a large fraction of the metals made by the system are now at distances $>R_{\rm vir}$. Among KBSS galaxies with unbound \ion{C}{4}, at least 20\% of the \ion{C}{4} column density is unbound. So it is plausible that a sizable fraction of the metals formed within the galaxy leave the halo. Indeed, KBSS galaxies show significant column densities of \ion{C}{4} and relatively high covering fractions at much larger distances (Rudie et al. in prep). \citet{gcr12} found significant enhancement in \ion{H}{1} to 300 pkpc. If much of this gas is also metal enriched, it could correspond to a significant reservoir of metals. Given that 70\% of galaxies in this sample with detected metals have some unbound metal-enriched gas, it appears likely that some fraction of the assembled metal mass lies outside of $R_{\rm vir}$. Future measurements from the KBSS sample will help to address the mass of metals beyond $R_{\rm vir}$.

\section{Summary}

In this paper, we have presented detailed measurements of the column densities, kinematics, and internal energy of metal-bearing gas within 100 pkpc ($\sim R_{\rm vir}$) of eight $L_{\rm UV}^*$ galaxy at $z\sim2$. These measurements are based on detailed modeling of high-resolution and high signal-to-noise ratio background QSO spectra which provide the first detailed view of the properties of gas within the halo of typical high-redshift galaxies. Our findings are as follows:

\begin{enumerate}
    \item The high-redshift circumgalactic medium is multi-phase. Singly, doubly, and triply ionized species typically share the same complex kinematic structure (Figure \ref{BX182} and Section \ref{fits}). However, an additional phase of gas is commonly detected in broad \ion{O}{6} and \ion{C}{4}. This gas is found at similar velocities with respect to the galaxy systemic redshifts as the narrow lower-ionization gas; however, its kinematic structure is characterized by much broader absorption features (Figure~\ref{b_CIV_OVI}), consistent with a possible hot, metal-enriched wind.
    
    \item The CGM gas is kinematically complex (Figure \ref{BX182}) with $>10$ absorption components commonly found per galaxy (Section \ref{cf_section} and Table \ref{components}). 

    \item The covering fraction of metal ions within 100 pkpc of high-z galaxies is high (Figures \ref{cf} and \ref{cf_sumN}, Section \ref{cf_section}). For column densities of $\sim10^{12}$ \cm2, the covering fraction for doubly and triply ionized species is 87-100\%. For \ion{C}{4}, the covering fraction is above 50\% for column densities $N_{\rm CIV} \geq 10^{13.5}$ \cm2. 
    
    \item \ion{O}{6} is challenging to measure in the high-$z$ CGM due to \ion{H}{1} contamination; however, when the relevant spectral range is free of contamination, we invariably find $N_{\rm OVI}>10^{13.9}$ \cm2 (Figures \ref{cf} and \ref{cf_sumN}, Section \ref{cf_OVI}). 
    
    \item 70\% of galaxies with detected metal absorbers have some gas that is unbound from the halo (Figure \ref{vesc}, Section \ref{kinematics}). Unbound absorbers are detected in all species (\ion{Si}{2} - \ion{O}{6}, Figure \ref{kinematic_hist}) suggesting this gas is multi-phase; however, the unbound gas is more commonly detected in high-ionization transitions. Further, unbound metals are not commonly associated with high \ion{H}{1} column densities, suggesting the absorbers are likely highly ionized, metal rich, or both. Given their high velocities and high levels of enrichment and/or ionization, a plausible origin for this material is from galactic winds. In this scenario, this gas represents metal-enriched material that would be permanently removed from the galaxy.
    
    \item The widths of absorbers of elements with different atomic masses constrain the temperature and non-thermal motions of gas within the CGM (Figure \ref{stack}, Section \ref{temp}). The typical inferred gas temperature within the CGM is $T = 10^{4.3-5.0}$~K depending on the sample used (Figure \ref{T_lim}). Roughly half of the detected gas has intermediate temperatures $10^{4.5} < T < 10^{5.5}$ K, hotter than the IGM, but cooler than the expected virial temperature of halos with $M_{\rm halo}=10^{12}~M_\odot$. Gas at these temperatures has short cooling times so that its persistence requires constant re-heating or replenishment of the warm gas phase (Section \ref{temp_discuss}). This gas may provide important clues about the nature of inflows and outflows at high redshift.
    
    \item Non-thermal motions in the halo gas are low ($ \langle  v_{\rm turb} \rangle < 6$ \kms, Figure \ref{turb_hist}, Section \ref{temp}) and are nearly always sub-dominant energetically (Figure \ref{energy_hist}). Thermal energy contributes $>90\%$ of the total internal energy in 58\% of the measured absorbers.
    
    \item Most of the detected gas within the CGM appears to be warmer than is typical of \ion{H}{1} absorbers within the IGM; however, there are few differences between the measured temperature distribution of metal-bearing absorbers in IGM samples \citep{rau96,kim16} and those CGM absorbers presented here, suggesting that many metal-rich absorbers in the high-$z$ IGM have recently been heated or ejected from a galaxy (Section \ref{IGM_discuss}). 
    
    \item Observations of one halo probed by two images of a background lensed QSO suggest the overall structure of halo gas that gives rise to metal absorbers has a physical size $>400$ pc. \ion{O}{6}-bearing gas within these structures appears to be homogeneous while \ion{C}{4} and lower-ionization gas exhibit variations in density and velocity structure on these scales (Section \ref{size} and Figure \ref{BX182_imageb}). This suggests that metal enriched zones within the halo likely have size scales of ~kpc or larger but with significant internal structure that varies on smaller scales. The \ion{O}{6}-bearing phase appears to be more volume filling than the lower-ionization gas. 
    
     \item The inferred mass of metals within the halo is large, equivalent to $25\%$ of the metal mass within the ISM (Section \ref{mass_discuss}). However, the estimated halo$+$ISM$+$stellar metal mass may account for less than 50\% of the mass of metals formed by the stellar populations, depending on the true values of the IMF-weighted yields, suggesting that many of the metals may reside in unprobed regions or phases of the CGM.
\end{enumerate}

This work represents the first detailed characterization of metal-enriched gas within the halos of high-redshift $L^*$ galaxies; however, much more information is needed. 

Given the measured temperatures of gas in the CGM of KBSS galaxies, we expect the densest and most metal-rich regions to have a rich
emission line spectrum from the many UV transitions that dominate the cooling of warm collisionally-ionized gas.  Ongoing observations with the
Keck Cosmic Web Imager \citep{KCWI} have sufficient sensitivity to detect such emission from KBSS galaxies.

Future analysis of the data presented in this work will address the ionization conditions and abundance patterns within the CGM, providing additional data on the nature of halo gas. The KBSS data also constrain the distribution of metal absorbers on larger scales and allow for the first detailed measurements of the evolution of the metal content of the CGM as a function of redshift. Forthcoming work will address these important issues.

\acknowledgments

We thank collaborators K. Adelberger, M. Bogosavljevi\'{c}, D. Erb, N. Konidaris,  D. Law,  O. Rakic, and M. Turner for their contributions to the KBSS survey. The authors thank S. Johnson, A. Benson, A. Piro, A. Newman, M. Rauch, H.-W. Chen, F. Zahedy, E. Schneider, D. Fielding, M. Gronke, J. Kollmeier, and S. Garrison-Kimmel for useful discussions which shaped the content or direction of this manuscript. We also wish to acknowledge the staff of the the W.M. Keck Observatory whose efforts ensure the telescopes and instruments perform reliably. Further, we extend our gratitude to those of Hawaiian ancestry on whose sacred mountain we are privileged to be guests. Early phases of this work were supported by NSF through grants AST-0908805 and AST-1313472. This research has made use of the Keck Observatory Archive (KOA), which is operated by the W. M. Keck Observatory and the NASA Exoplanet Science Institute (NExScI), under contract with the National Aeronautics and Space Administration.

 \facility{Keck (LRIS)}, \facility{Keck (MOSFIRE)}, \facility{Keck (HIRES)}
\software{VPFIT, Grackle \citep{smi17}}

\bibliographystyle{yahapj}
\bibliography{references}

\end{document}